\documentclass[aps,pra,epsf,superscriptaddress,amsmath,amssymb,amsfonts,twocolumn,showpacs,nofootinbib]{revtex4-1}
\usepackage{overpic}
\usepackage{graphicx}
\usepackage{epsfig}
\usepackage{dcolumn}
\usepackage{bm}
\usepackage{braket}
\usepackage{amsmath}
\usepackage{mathtools}
\usepackage{graphicx,color,xcolor}
\usepackage{hyperref}
\newcommand{\abs}[1]{\left| #1 \right|}
\usepackage{hyperref}
\usepackage{multirow}
\usepackage{tikz}
\usetikzlibrary{positioning,shapes}

\usepackage[normalem]{ulem} 

\newcounter{fig}

\newcommand{\ii}{\mathrm{i}}
\DeclareMathOperator*{\dprime}{\prime \prime}

\newcommand\mybar{\kern1pt\rule[-\dp\strutbox]{.8pt}{\baselineskip}\kern1pt}

\usepackage[breakable, theorems, skins]{tcolorbox}
\tcbset{enhanced}

\begin{document}

\title{Two-component droplet phases and their {spectral} stability in
one dimension}
\author{E.~G.~Charalampidis}
\email{echaralampidis@sdsu.edu}
\affiliation{Department of Mathematics and Statistics and Computational
Science Research Center, San Diego State University, San Diego, CA 92182-7720, USA}
\author{S.~I.~Mistakidis}
\email{smystakidis@mst.edu}
\affiliation{Department of Physics, Missouri University of Science and
Technology, Rolla, MO 65409, USA}
\date{\today}

\begin{abstract}

We unravel the existence and stability properties of one-dimensional droplets arising in
genuine two-component particle imbalanced bosonic mixtures under the influence of a weak
harmonic confinement.~A plethora of miscible droplet phases is found with the majority
component atoms in the vicinity of the minority ones assembling in a droplet configuration
while the remaining excess atoms being in the gas state.~Our computations identify a transition from
a bound to a trapped gas many-body state for increasing total atom number and fixed interactions,
with the bound character being prolonged for stronger intercomponent attractions.~Spectral
stability of these two-component droplets is revealed by examining their underlying
Bogoliubov-de-Gennes excitation spectrum.~Furthermore, the collective dynamics associated with
the fundamental dipole and breathing modes of the system is monitored by suddenly shifting the
trap's center or quenching the trap's frequency, respectively.{~There, a back action of the significantly
localized minority component is imprinted in the majority cloud.~The present findings can be simulated
in recent cold-atom experiments, thus setting the stage for probing the complex nonequilibrium
dynamics of two-component droplets.}

\end{abstract}

\maketitle

\section{Introduction}

Quantum droplets are ultradilute and incompressible many-body states of matter arising at
attractive interparticle interactions in three-dimensions, and are sustained due to the
presence of quantum fluctuations~\cite{Luo2020,bottcher2020new,chomaz2022dipolar,mistakidis2023few}.%
~They have been experimentally observed with the aid of recent cold atom simulators originally
utilizing dipolar Bose gases~\cite{Schmitt2016,Chomaz2016} and short-range bosonic mixtures
afterwards~\cite{Cabrera2018,Cheiney2018,Semeghini_drops,D'Errico,Guo_heteron}, thus providing
access to beyond mean-field phenomena.~According to current investigations, the latter may be
predominantly modeled with the dimension~\cite{Zin_crossover,pelayo2024phases} and effective
range~\cite{schutzhold2006mean,Petrov_stabilization} dependent first-order Lee-Huang-Yang
(LHY)~\cite{properties2005eigenvalues} quantum correction term.~This contribution can be consistently
incorporated into an extended Gross-Pitaevskii (eGPE)~\cite{Petrov_stabilization,Petrov_lower_dim}
framework, {being itself} suitable for describing not only droplets, but also other bound states such as
bubbles~\cite{katsimiga2023interactions,Edmonds_dark_drops} and supersolids~\cite{Blakie_SS,Bland_SS}.%
~Restricting ourselves to short-range bosonic mixtures that we will study herein, a variety of
droplet properties have been explored.~These refer, for instance, to ground state droplet
configurations~\cite{Astrakharchik_1Ddrops}, their excitation spectrum~\cite{Tylutki,Hu_excitation,Sturmer_breath},
their potential inelastic  collisions~\cite{Astrakharchik_1Ddrops,Ferioli_collision}, their
coexistence with nonlinear excitations~\cite{Li_vortex,Katsimiga_sol_drops,kartashov2022spinor},
rotational properties~\cite{Tengstrand_rot,Yoifmmode,Liangwei}, and modulation instability
events~\cite{mithun2020modulational,otajonov2022modulational} attributed to their attractive nature.

{However, the main focus of the aforementioned studies has been placed on
droplet states whose components density
ratio is proportional to their intracomponent coupling one resulting in energetically
stable structures~\cite{Petrov_stabilization}.~Hence, two-component droplet states violating this condition are far
less explored.~In this context, intercomponent atom imbalance is anticipated to result in: i) energetically
lower and less stable droplets as compared to the previous case, and ii) allow the formation of enriched phases including mixed
configurations. For the latter, depending on the interactions, a portion of the majority component atoms cannot bind
to the droplet anymore~\cite{Flynn_box,Flynn_trap_drops}.~Recent investigations explored, to some extent,
relevant ground states of droplets in three dimensions (3D), revealing the coexistence of droplet and gas
phases, {and} analyzed the interplay of the behavior of their collective modes and self
evaporation~\cite{Flynn_box,Flynn_trap_drops}.}


{In this sense,} understanding the existence of highly {particle/interaction} imbalanced two-component droplet states
in one-dimension (1D), {and more crucially}, their spectral stability is still far from
complete~\cite{Englezos_imb}.~Notable examples here, include rotational droplet properties in a ring
geometry~\cite{Tengstrand}, lattice trapped droplets building upon a super Tonks-Girardeau
gas~\cite{valles2024quantum}, the existence of multipole droplet states~\cite{Kartashov_multipoles},
and beyond-LHY approaches to explicate the underlying correlation patterns~\cite{Englezos_imb,mistakidis2023few}.%
{~Along these lines, the interplay of the distinct components in 1D remains elusive.  It holds
the premise of giving rise to previously unexplored mixed droplet-gas configurations, while unveiling,
among others, their potential miscibility and back action properties~\cite{Pelayo_BF_drops} together
with their spectral stability and dynamical response.}%
~We make a first step in this direction herein {too} by monitoring the dynamics of certain
collective modes.

To examine the aforementioned possibilities we deploy an 1D two-component bosonic gas characterized
by intracomponent repulsion and intercomponent attraction while featuring high population imbalance.%
~Recall that 1D settings are promising for studying droplet formation due to their lower
densities~\cite{Lavoine,Tolra_1D}.~Indeed, the latter act against lossy mechanisms, and thus self-evaporation
being a central issue for the long-time observation of these structures in higher
dimensions~\cite{fort2021self,Cabrera2018}.~{We describe the static properties and spectral stability,
as well as dynamical response of the two-component droplet setting subjected to weak harmonic confinement
via the appropriate set of coupled eGPEs incorporating the 
LHY beyond mean-field 
correction
term~\cite{Petrov_lower_dim,mithun2020modulational}}.

At first, we showcase the existence of distinct miscible two-component particle-imbalanced
stationary droplet solutions upon varying the total particle number as well as both
intra- and inter-component interactions.~Composite configurations are identified with
the minority component forming flat-top droplets, whilst the majority one hosts a
droplet fragment overlapping with the minority one but existing on top of a finite
background whose atoms are in the gas phase.~Strikingly, the minority component and
the droplet fragment of the majority are separated by a sharp ``domain-wall" boundary
for strong attractions.

{These 1D states studied herein bear similarities with the 3D ones reported
in~\cite{Flynn_trap_drops,Flynn_box}, and expose the tunability of the ensuing
configurations.~Indeed, we explicate that the fraction of these individual
fragments can be tuned through parametric variations of: i) the involved
interactions, ii) particle imbalance, and iii) atom number.~We demonstrate
that as the attraction gets stronger, it facilitates the decrease of the
majority's droplet segment.~Controllable transition from bound state droplets
(negative chemical potential) to a trapped gas phase (positive chemical potential)
is achieved by tuning the atom number and interactions.~Particularly, it is found
that the interval of existence of bound states is enhanced for increasing number
of atoms predominantly for stronger attractions, and to a lesser extent for increasing
imbalance.}

{A core finding of the present work is that the identified two-component droplet
states are spectrally stable over the atom number and interactions}.~This is achieved
by numerically extracting the underlying excitation spectrum through a Bogoliubov-de-Gennes
(BdG) linearization analysis, thus generalizing the stability of both
1D~\cite{Tylutki,katsimiga2023interactions} and two-dimensional (2D)~\cite{Li_2Dvortex_drop,bougas2024stability}
symmetric droplet states.~{Moreover, the lowest-lying  collective modes, i.e.,
dipole and breathing, of the resultant spectrum are identified and dynamically probed by applying suitable trap quenches}.~Namely, we show
that the dipole mode frequency coincides with that of the trap independently of
parametric variations, whilst the breathing mode frequency decreases slightly for
larger atom numbers.~The droplet and the excess atom fragments of the majority
component oscillate in-sync, while a density hump builds atop the majority species
distribution as a result of the minority species back action.

This work is organized as follows.~Section~\ref{sec:EGPE}, introduces the attractively interacting two-component bosonic system utilized to enter the droplet regime, and the ensuing coupled set of eGPEs used to describe the stationary configurations and nonequilibrium dynamics of two-component droplets.~In Section~\ref{sec:statics} we analyze the characteristics of the stationary two-component droplet configurations along with their stability properties. Section~\ref{sec:dynamics} explicates the collective breathing and dipole mode dynamics of the two-component droplets after appropriate quenches of their external traps are applied.~We summarize and elaborate on future research directions in Section~\ref{sec:conclusions}. 
{Appendix~\ref{sec:Appendix1} provides details on the matrix elements of the underlying eigenvalue problem.}

\section{Two-component Bose droplet}\label{sec:EGPE}

We employ an 1D weakly trapped pseudo-spin-$1/2$ bosonic system, i.e., $m_1=m_2 \equiv m$,
experiencing particle imbalance ($N_1 \neq N_2$) and lying in the droplet regime.~The latter
is entered in the case of intracomponent repulsion ($g_1 \neq g_2 >0$) and intercomponent attraction
($g_{12}<0$) of the involved spin states but also in the presence of quantum fluctuations that
facilitate the generation of droplet configurations~\cite{Petrov_lower_dim,Luo2020,mistakidis2023few}.%
~The 1D nature of this setting is warranted due to the weak harmonic confinement that is elongated
in the $x$-direction with {dimensionless} frequency $\Omega=10^{-3}$ {(see also below for the units)},
{along with} the strong transverse confinement
($\Omega_{\perp} \gg \Omega$). 
{This condition essentially prevents structure formation in these
directions~\cite{exp_lower_dim,Romero_Per}.}
~In a corresponding experiment, our system may be realized
through two distinct hyperfine states of $^{39}$K as in the 3D experiments of Refs.~\cite{Cabrera2018,Semeghini_drops}.

Specifically, 1D two-component quantum droplets in the presence of
the first-order LHY quantum correction and a parabolic external potential
are described, in the weakly interacting regime, by the following dimensionless coupled eGPEs~\cite{mithun2020modulational}
{\small
\begin{subequations}\label{eq:eGPE}
\begin{align}
\ii \frac{\partial \Psi_1(x,t)}{\partial t}&\!=\!\Bigg [\!-\! \frac{1}{2}\frac{\partial^2 }{\partial x^2} + (P+GP^{-1}) |\Psi_1|^2 %
-(1 -G) |\Psi_2|^2  \notag \\
&-\!\frac{P}{\pi} \sqrt{P|\Psi_1|^2 +P^{-1}|\Psi_2|^2 } + V(x) \Bigg ] \Psi_1(x,t), \label{eq:eGPEa} \\
&\notag\\
\ii \frac{\partial \Psi_2(x,t)}{\partial t}&\!=\!\Bigg [\!-\! \frac{1}{2}\frac{\partial^2 }{\partial x^2} + (P^{-1}+GP) |\Psi_2|^2 %
-(1 -G) |\Psi_1|^2 \notag \\ %
&-\!\frac{1}{P \pi} \sqrt{P^{-1}|\Psi_2|^2 +P|\Psi_1|^2 } + V(x) \Bigg ] \Psi_2(x,t). \label{eq:eGPEb}
\end{align}
\end{subequations}
}
In these expressions, $\Psi_{j}(x,t)$ (with $j=1,2$) correspond to the 1D wave function
of the $j$th component, and $V(x)=\Omega^{2}x^{2}/2$ models the external harmonic trap of
strength $\Omega$. We remark that an arguably weak trap is used such that the validity of
the LHY contribution (holding within the local density approximation~\cite{Petrov_stabilization,Petrov_lower_dim})
is ensured.~The impact of the external trap on the LHY form remains still an open issue for
future research.~Moreover, $G=2g\delta g/(g_1+g_2)^2$, $P=\sqrt{g_1/g_2}$ with $g=\sqrt{g_1g_2}$
representing the mean average intracomponent repulsion while $\delta g = g + g_{\rm 12}$ is the
mean-field balance point~\cite{Petrov_stabilization,mistakidis2023few}.~In the following, we
measure the length, time and wave function in terms of $\xi$, $\hbar/(m \xi^2)$, and
$\frac{\sqrt{g_1}+\sqrt{g_2}}{\sqrt{\pi \xi}(2\abs{\delta g})^{2/3}}$, where
$\xi=\frac{\hbar^2 \pi}{m} \frac{\sqrt{2 \abs{\delta g}}}{ g(\sqrt{g_1}+\sqrt{g_2})}$ denotes
the healing length~\cite{Tylutki}.~It is interesting to note that the above two-component system
can be reduced to an effective single-component one {associated with the term symmetric droplet below}.
{This may happen by requiring either $g_1=g_2 \equiv g$, $N_1 = N_2 \equiv  N$ and $\Psi_1=\Psi_2 \equiv \Psi$ or $\abs{\Psi_1}^2\sqrt{g_2} = \abs{\Psi_2}^2\sqrt{g_1}$.}
~This single
droplet reduction has been extensively studied, see, e.g., the reviews~\cite{Luo2020,mistakidis2023few},
and it arguably hosts less complex configurations than the ones presented next.

\section{Existence and Stability of Particle Imbalanced
Droplets}\label{sec:statics}

Below, we explore the existence of stationary two-component droplet configurations for
different atom numbers and intercomponent interactions, while mainly keeping fixed the
intracomponent couplings.~Subsequently, we study the excitation spectrum of these
configurations in order to infer their stability properties.

\subsection{Two-component droplet distributions}\label{sec:results_exstab}

To visualize the emergent two-component droplet configurations we employ the density
distributions of each component, namely $\abs{\Psi_1(x)}^2$ and $\abs{\Psi_2(x)}^2$.%
~These are obtained as follows.~We identify stationary states of the system of
Eqs.~\eqref{eq:eGPEa}-\eqref{eq:eGPEb} by employing the ansatz
\begin{align}\label{eq:stat_ansatz}
\Psi_{j}(x,t)=\psi_{j}^{(0)}(x)e^{-\ii \mu_{j} t},
\end{align}
where $\psi_{j}^{(0)}$ and $\mu_{j}$ stand for the stationary wave function and the 
chemical potential of the $j$th component, respectively.~Substitution of Eq.~\eqref{eq:stat_ansatz}
into Eqs.~\eqref{eq:eGPE} leads to the following time-independent eGPE system
\begin{subequations}
\begin{align}
\mu_{1}\,\psi_{1}^{(0)}&=-\frac{1}{2}\psi_{1}^{\dprime(0)}+\Bigg [(P+GP^{-1}) |\psi_1^{(0)}|^2-(1 -G) |\psi_2^{(0)}|^2\notag \\
&-\frac{P}{\pi} \sqrt{P|\psi_1^{(0)}|^2 +P^{-1}|\psi_2|^2 } + V(x)\Bigg ]\psi_{1}^{(0)},
\label{eq:stat_eGPEa}\\
&\notag\\
\mu_{2}\,\psi_{2}^{(0)}&=-\frac{1}{2}\psi_{2}^{\dprime(0)}+\Bigg [(P^{-1}+GP) |\psi_2^{(0)}|^2-(1 -G) |\psi_1^{(0)}|^2\notag \\
&-\frac{1}{P \pi} \sqrt{P^{-1}|\psi_2^{(0)}|^2 +P|\psi_1^{(0)}|^2 } + V(x) \Bigg ] \psi_2^{(0)}(x,t).
\label{eq:stat_eGPEb}
\end{align}
\end{subequations}
Then, we solve Eqs.~\eqref{eq:stat_eGPEa}-\eqref{eq:stat_eGPEb} numerically by using Newton's
method~\cite{kelley_book1,kelley_book2}.~Specifically, we pose the above system on an 1D computational
grid with half-width $L$ (here, $L=2000$), and discretize derivatives with a centered,
fourth-order accurate finite difference scheme with resolution $dx=0.2$ while zero Dirichlet boundary
conditions are imposed at the edges of the grid.~Branches of numerically exact droplet solutions are
obtained by utilizing the following strategy. 

\begin{figure*}[pt!]
\begin{center}
\begin{overpic}[height=0.32\textheight, angle =0]{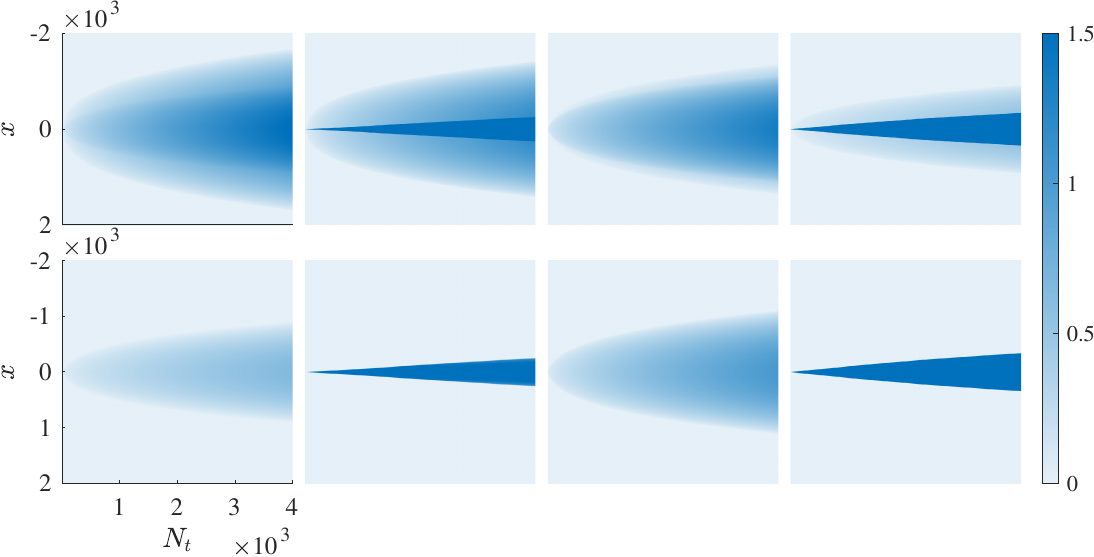}
\put(6.5,45){$\textrm{(a)}$}
\put(28.7,45){$\textrm{(b)}$}
\put(51,45){$\textrm{(c)}$}
\put(73.2,45){$\textrm{(d)}$}
\put(6.5,24){$\textrm{(e)}$}
\put(28.5,24){$\textrm{(f)}$}
\put(51,24){$\textrm{(g)}$}
\put(73.2,24){$\textrm{(h)}$}
\end{overpic}
\end{center}
\caption{
{
\label{fig1_summary}
(Color online)
Stationary density distributions of (a)-(d) the majority ($|\psi_{1}|^2$), and
(e)-(f) the minority ($|\psi_{2}|^2$) droplet configurations for $g_{1}=g_{2}=1$,
with intracomponent interactions of (a), (e), (c), (g) $g_{12}=-0.1$, and
(b), (f), (d), (h) $g_{12}=-0.8$, as a function of the total
particle number $N_t$.~The two-component bosonic system features
($N_{1}=80\% N_t$, $N_{2}=20\% N_t$) and ($N_{1}=60\% N_{t}$, $N_{2}=40\% N_{t}$)
particle imbalances in (a)-(b), (e)-(f), and (c)-(d), (g)-(h),
respectively, whilst it is subject to a weak harmonic trap of frequency $\Omega=10^{-3}$.%
~An increasing atom number results in a larger amount of bound majority atoms to the
minority ones and a broader distribution of excess particles in the majority.~Naturally,
this process is more pronounced for stronger intercomponent attractions, i.e., $g_{12}=-0.8$,
see panels (b) and (f), and (d) and (h).~On the other hand, for smaller attractions, i.e.,
$g_{12}=-0.1$ the distribution of each component is Gaussian-type, and the segment of excess
atoms is suppressed, see panels (a) and (e), and (c) and (g).~All quantities shown are in
dimensionless units.
}
}
\end{figure*}

First, the system of Eqs.~\eqref{eq:stat_eGPEa}-\eqref{eq:stat_eGPEb} is reduced to its
single-component counterpart~\cite{Petrov_lower_dim} by assuming  $\psi_{1}=\psi_{2}=\psi$,
$\mu_{1}=\mu_{2}=\mu$, and $g_{1}=g_{2}=g$.~Then, the exact 1D droplet solution of%
~\cite{Astrakharchik_1Ddrops} is employed, and fed to our nonlinear solvers as an initial guess.%
~Upon tuning the involved interaction coefficients ($g_{1}$, $g_{2}$, $g_{12}$) through numerical continuation~\cite{kuznetsov_book}, we reach specific symmetric droplet configurations having $N_1=N_2$.
Here, the number of atoms in each component is given by $N_{j}=\int_{-L}^{L}|\psi_{j}|^{2}\,dx$.%
~Consecutively, these solutions are used to determine imbalanced ones by considering parametric
variations of either the total particle number $N_{t}=c\,N_{1}+\left(1-c\right)\,N_{2}$ or the
imbalance fraction $c\in[0,1]$, again, through continuation~\cite{kuznetsov_book}.~In both cases,
the values of the chemical potentials $\mu_{j}$ are obtained in the course of the continuation by
adding the algebraic equations $N_{j}^{\mathrm{target}}-\int_{-L}^{L}|\psi_{j}|^{2}\,dx=0$ to
the nonlinear system of Eqs.~\eqref{eq:stat_eGPEa}-\eqref{eq:stat_eGPEb}, where $N_{j}^{\mathrm{target}}$
is a ``target'' value for the $j$th atom number.~In the following, the characteristics of the
two-component droplet configurations having fixed trap frequency $\Omega=10^{-3}$ for high and
small particle imbalances, i.e.,~($N_1=80\%N_t$, $N_2=20\% N_t$) and ($N_1=60\%N_t$, $N_2=40\% N_t$),
respectively, are analyzed in Fig.~\ref{fig1_summary} (see, the caption therein for system's parameter
values).

\begin{figure*}[!pt]
\begin{center}
\includegraphics[height=.19\textheight, angle =0]{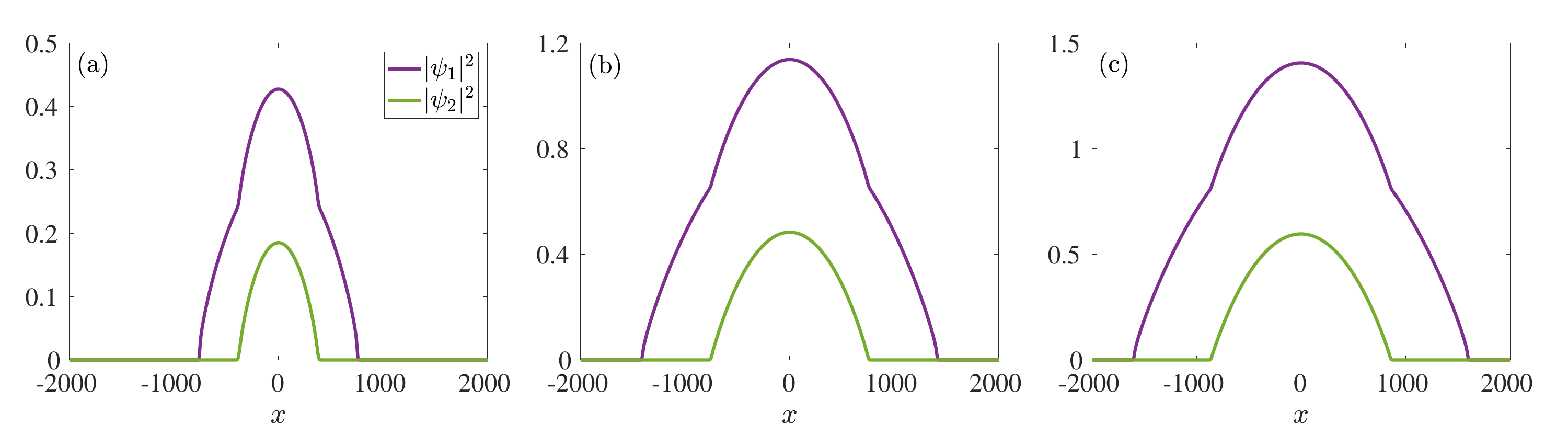}\\
\begin{tikzpicture}
\node(a){\includegraphics[height=.195\textheight, angle =0]{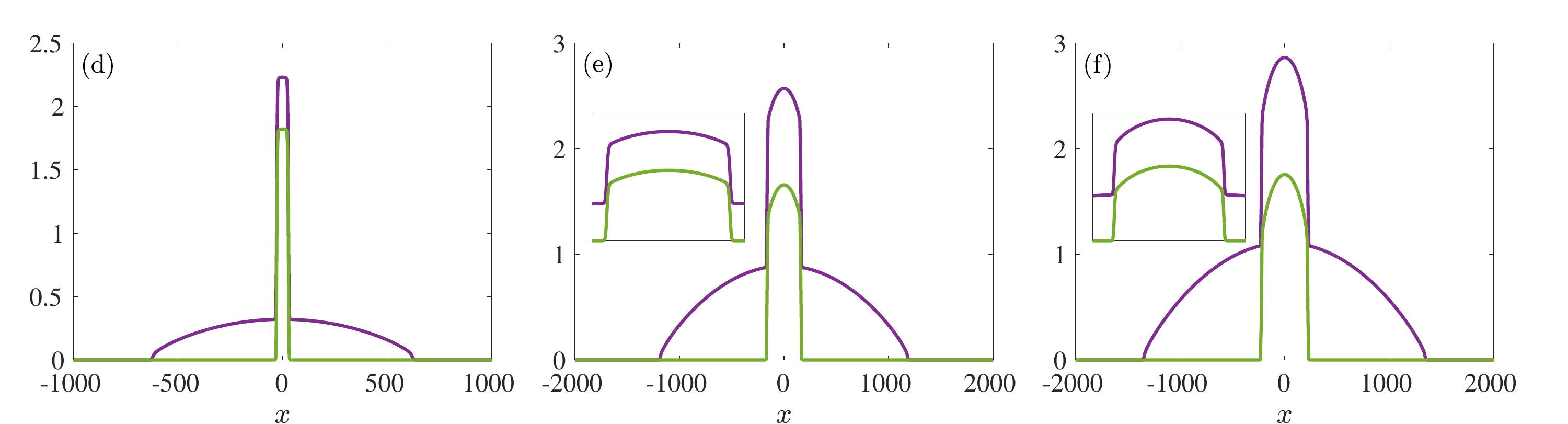}};
\node at(a.center)[draw, black, line width=1pt,dashed, ellipse, minimum width=22pt, minimum height=2pt,rotate=-90,yshift=-17pt,xshift=-6pt]{};
\node at(a.center)[draw, black, line width=1pt,dashed, ellipse, minimum width=25pt, minimum height=2pt,rotate=-90,yshift=136pt,xshift=-7pt]{};
\end{tikzpicture}
\end{center}
\caption{
\label{fig:den_profs}
(Color online)
Stationary density profiles of each component (see legends) of the particle imbalanced
$N_1=80\%N_t$, $N_2=20\%N_t$ bosonic setting for (a), (d) $N_t=500$, (b), (e) $N_t=2500$
and (c), (f) $N_t=3500$.~The harmonically trapped two-component system with $\Omega=10^{-3}$
features $g_1=g_2=1$ and (a)-(c) $g_{12}=-0.1$ and (d)-(f) $g_{12}=-0.8$.~The configurations
show a central droplet component and a fraction of excess atoms in the majority component.%
~Insets of panels (e), (f) provide a magnification of the densities at the droplet core
showcasing its flat-top shape, {while the dashed black ellipses therein highlight the sharp
interface formed between the minority and majority distributions}.~The amount of
bound atoms of the majority component increases for smaller attractions or larger $N_t$.~Dimensionless
units are utilized for the different parameters.}
\end{figure*}

The density distributions of the two-component droplet setting featuring
($N_1=80\%N_t$, $N_2=20\% N_t$) imbalance are depicted in Figs.~\ref{fig1_summary}(a),(e)
and (b),(f) with respect to the total atom number for weak and strong intercomponent
attractions, respectively.~As it can readily seen, the densities of both the majority
and the minority components become wider for increasing atom number, and irrespectively
of the intercomponent attraction, see also exemplary density profiles in
Fig.~\ref{fig:den_profs}.~The configurations remain always miscible because the minority
atoms reside within the majority cloud while binding a portion of the latter in their
vicinity.~Interestingly, the remaining atoms of the majority component are excess particles,
and they do not participate in the droplet part of the mixture.~They are rather in a gas
state as argued for the corresponding 3D system~\cite{Flynn_box}, and contribute negatively
to the bound state nature of the total system.~This will be shown below by evaluating the
involved chemical potentials (see, Fig.~\ref{fig:chem_pot}).

The bound and un-bound portions can be easily discerned in the substantially deformed
density distributions especially of the majority component where the central highly
localized segment refers to the bound fragment, and the spatially extended tail to
the excess atoms as illustrated, for instance, in Figs.~\ref{fig1_summary} and
Figs.~\ref{fig:den_profs}.~{Few-body analogues of these structures were discussed in Ref.~\cite{Englezos_imb}.}~Such mixed configurations, possessing both a droplet portion
and excess particles, appear to be the energetically lower many-body states of the
system, and rise exclusively in the presence of intercomponent imbalance.%
~Indeed, they cannot form in the symmetric droplet case, see, e.g.,
Refs.~\cite{Astrakharchik_1Ddrops,Tylutki}.~It is also worth noting that the structural
deformation of the majority component configurations is reminiscent of anti-dark states,
i.e., density humps on top of a finite matter-wave background, which can be generated in
repulsive Bose gases~\cite{Qu_magnetic,Katsimiga_AD}.~Moreover, this binding mechanism
among the majority and the minority atoms is more pronounced for either fixed interactions
and larger atom number, see Figs.~\ref{fig:den_profs}(d)-(f), or for stronger intercomponent
attractions and constant particle imbalance, upon comparing, in particular,
Figs.~\ref{fig:den_profs}(b) and (e).

It is also important to comprehend the explicit gradual deformation of the density
distributions of both components for fixed interactions and increasing number of atoms.~A
``knee" shape pattern starts to appear for relatively small $N_t$ upon the majority component
density around the edges of the minority cloud reflecting the presence of attractive intercomponent
interactions.~As $N_t$ increases, these patterns become more prominent, and a larger number
of atoms accumulates  towards the minority component and bind to it.

This phenomenon can be easily discerned,  especially at strong attractions, in the density
distributions depicted in Figs.~\ref{fig1_summary}(b) and (f), and the profiles illustrated
in Fig.~\ref{fig:den_profs}.~Notice also that the minority component tends to form a flat-top
configuration, see the insets of Fig.~\ref{fig:den_profs}(e), (f),  manifesting its incompressibility,
a characteristic of droplets~\cite{Petrov_stabilization,Gangwar_2024}.~The same flat-top signatures
occur for the fragment of the majority component distribution which is concentrated around the minority.%
~Moreover, the interface among the minority and the majority distributions is sharp, {see also the regions encircled by the  highlighted
dashed black ellipses in the insets of Figs.~\ref{fig:den_profs}(e)-(f).~This sharp boundary emanates from the ensuing droplets (of the majority segment and the minority species) whose tails are abrupt due to incompressibility.~Such interfaces
[cf.~Figs.~\ref{fig:den_profs}(d)-(f)] are strongly reminiscent of a ``domain-wall"
and are particularly interesting for dynamical applications of these structures.}

~Simultaneously, the fragment of the excess particles grows, and gets manifested by the
increasingly extended tails of the majority density distribution.~Eventually, we reach a
total atom number above which the majority component atoms cannot anymore bind to the
droplet segment, and the portion of the excess atoms increases, thus resulting in highly
extended density tails.~This increasing tendency of the segment of the excess particles is
arguably more prominent for stronger intercomponent attractions, e.g., compare
Figs.~\ref{fig:den_profs}(c) and (f).

Signatures of the aforementioned two-component, structurally deformed droplet phases occur
also for decreasing particle imbalances until eventually they disappear when the balance
limit is approached and the single droplet is attained.{~Examples of low imbalance
($N_{1}=60\% N$, $N_{2}=40\% N$) configurations close to the symmetric system are presented
in Figs.~\ref{fig1_summary}(c), (g) and (d), (h), for both weak and strong intercomponent
attractions, respectively.}~We observe that in the case of weak attractions such as
$g_{12}=-0.1$ [cf.~Figs.~\ref{fig1_summary}(c) and (g)], both components exhibit a
Gaussian-type distribution which becomes wider for larger $N_t$.~Here, the segment of
excess atoms is suppressed and is anticipated to completely vanish for a $N_{1}=N_{2}=50\% N$
mixture.{~However, an increasing attraction, e.g., to $g_{12}=-0.8$ enhances the mixed
character of the ensuing phases, and in particular, the droplet and gas fragments building
upon the majority component become noticeable, see Figs.~\ref{fig1_summary}(d) and (h).%
~Of course, they are not comparable to the high particle imbalance setup [cf.~Figs.~\ref{fig1_summary}(a), (e)
and (b), (f)].~Simultaneously, the minority component is significantly more localized than
the $g_{12}=-0.1$ case in order to sustain the droplet portion of the many-body state.}

\subsection{Bound state  to trapped gas transition}\label{chemical_pot}

\begin{figure*}[!pt]
\begin{center}
\begin{overpic}[height=.20\textheight, angle =0]{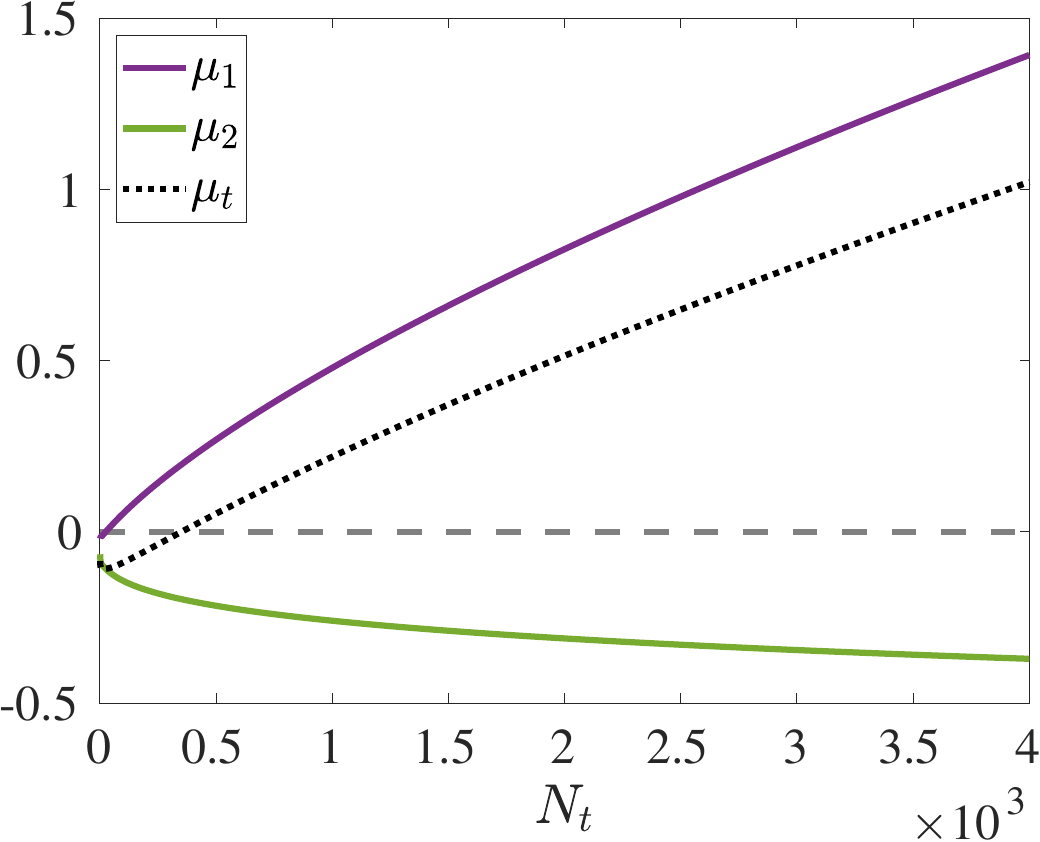}
\put(35,70){$\textrm{(a)}$}
\put(25.6,82){$g_{1}=g_{2}=1,\,g_{12}=-0.1$}
\end{overpic}
\begin{overpic}[height=.20\textheight, angle =0]{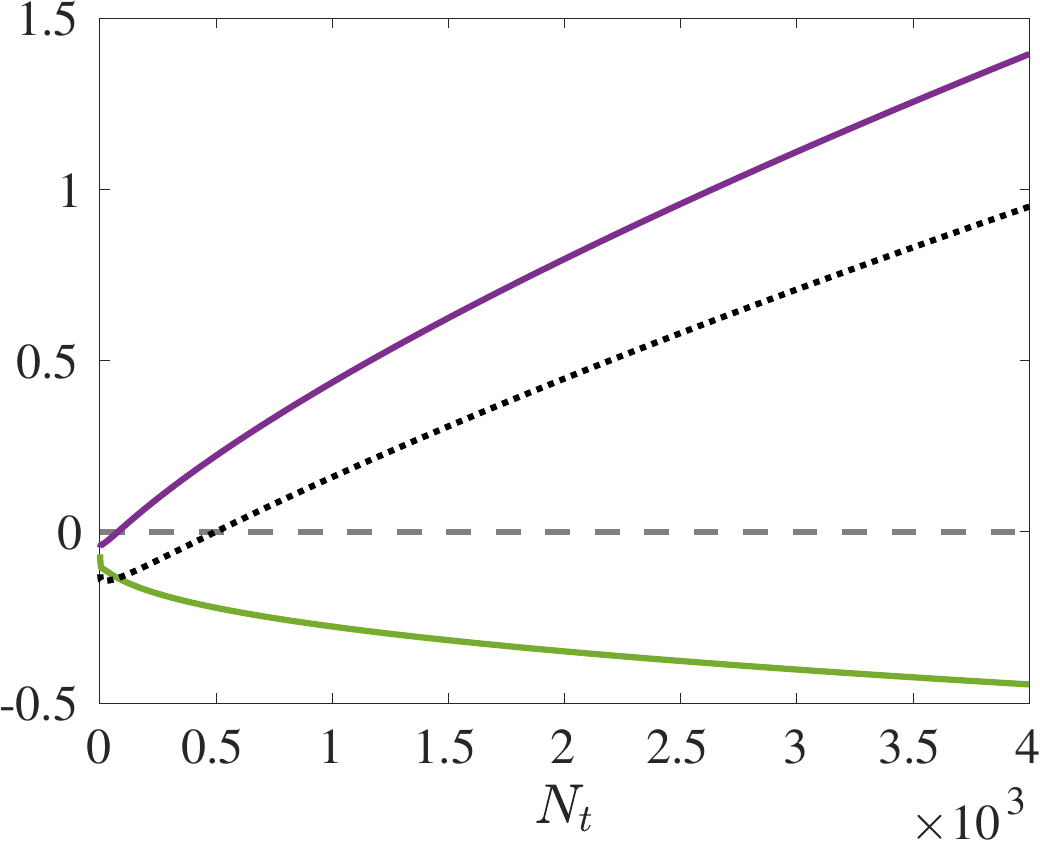}
\put(15,70){$\textrm{(b)}$}
\put(20.5,82){$g_{1}=1,\,g_{2}=0.5,\,g_{12}=-0.1$}
\end{overpic}
\begin{overpic}[height=.20\textheight, angle =0]{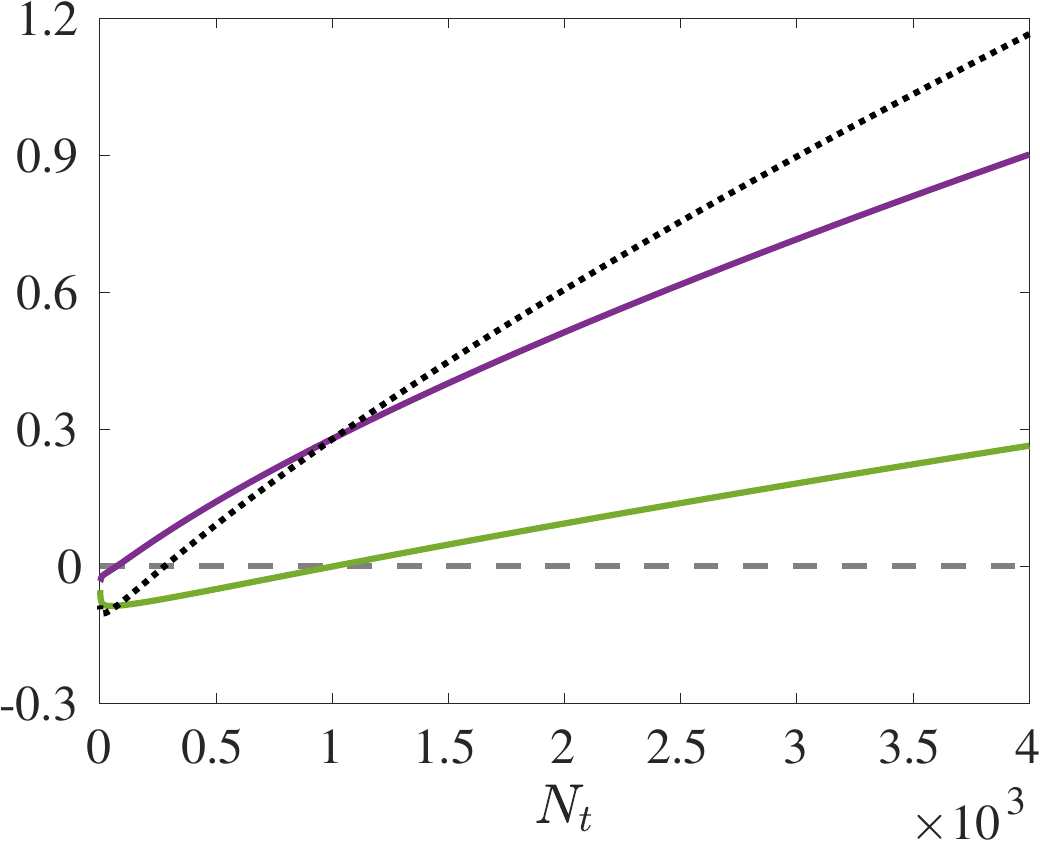}
\put(15,70){$\textrm{(c)}$}
\put(25.6,82){$g_{1}=g_{2}=1,\,g_{12}=-0.1$}
\end{overpic}
\vskip 1.0cm
\begin{overpic}[height=.20\textheight, angle =0]{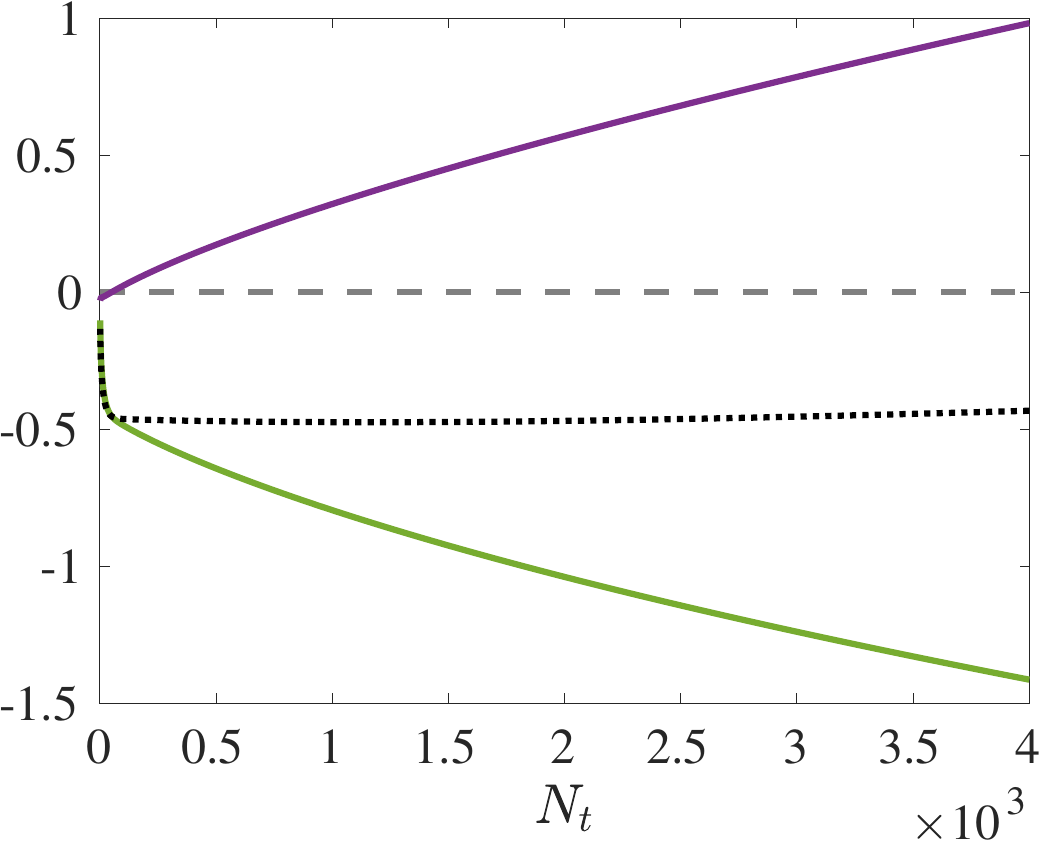}
\put(15,70){$\textrm{(d)}$}
\put(25.6,82){$g_{1}=g_{2}=1,\,g_{12}=-0.8$}
\end{overpic}
\begin{overpic}[height=.20\textheight, angle =0]{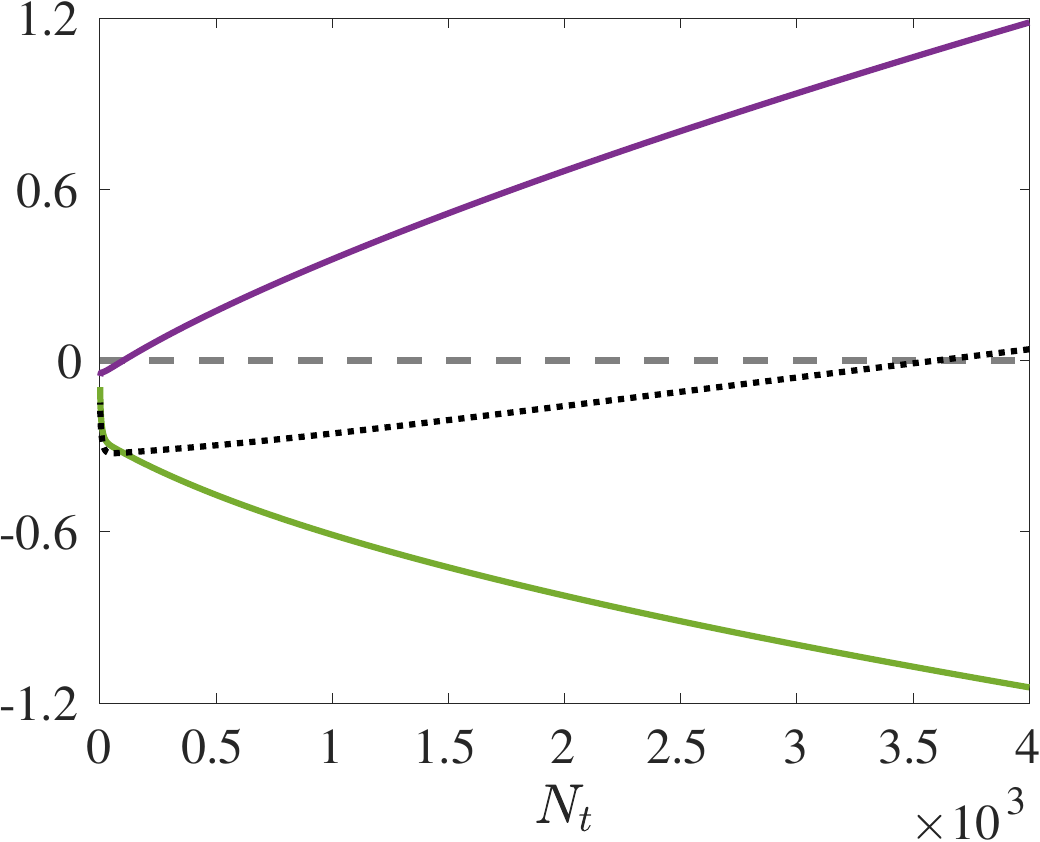}
\put(15,70){$\textrm{(e)}$}
\put(20.5,82){$g_{1}=1,\,g_{2}=0.5,\,g_{12}=-0.5$}
\end{overpic}
\begin{overpic}[height=.20\textheight, angle =0]{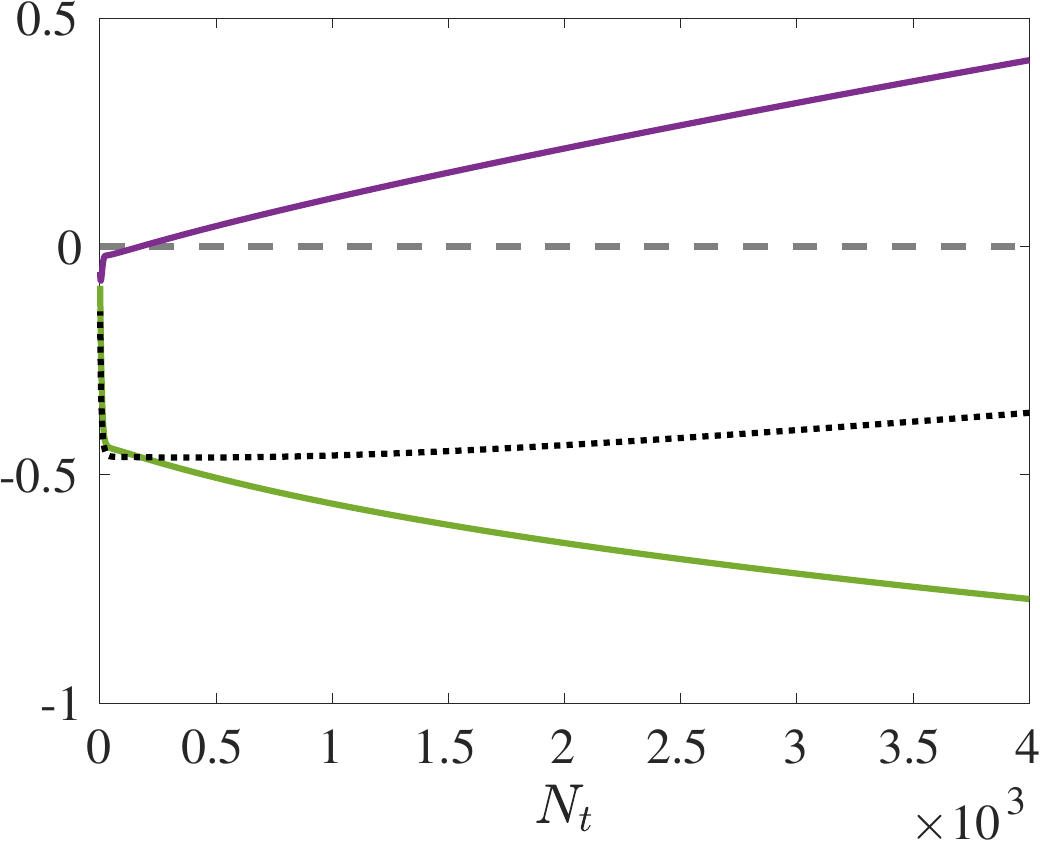}
\put(15,70){$\textrm{(f)}$}
\put(25.6,82){$g_{1}=g_{2}=1,\,g_{12}=-0.8$}
\end{overpic}
\end{center}
\caption{
\label{fig:chem_pot}
(Color online)
Chemical potential of each component along with the total chemical potential
(see the legend in panel (a)) of the particle imbalanced bosonic mixture with
respect to the total number of atoms.~The two-component droplet setting is characterized
by either a $N_1=80\%N_t$ and $N_2=20\%N_t$ [panels (a), (b), (d), and (e)] or a
$N_1=60\%N_t$ and $N_2=40\%N_t$ [panels (c) and (f)] imbalance. {The
ensuing interactions of the different settings are depicted above of each panel.}%
~The horizontal dashed grey line marks the position
of the zero chemical potential.~It is evident that at stronger intercomponent attractions
(lower panels) the system remains bound having negative total chemical potential.~This
emanates from the competition among the  minority and the majority components featuring
negative and positive chemical potentials, respectively.~However, a transition from bound
to a trapped gas state occurs for smaller attractions, see upper panels.~In all cases, an
external harmonic trap is present with frequency $\Omega=10^{-3}$, and the depicted quantities
are in dimensionless units.}
\end{figure*}

To understand the bound state character of the above-described two-component configurations,
we next evaluate not only each of the chemical potentials $\mu_j$, but also the total chemical
potential, $\mu_t=\mu_1+\mu_2$ of the system.~Naturally, droplet configurations occur for
negative chemical potential, and thus negative energy, since they are self-bound
states~\cite{bottcher2020new,chomaz2022dipolar}; otherwise a positive chemical potential
implies that a gas state takes place~\cite{pethick2008bose}.~These chemical potentials are
presented in Fig.~\ref{fig:chem_pot} for different yet fixed interaction configurations
($g_1$, $g_2$, $g_{12}$) and varying total atom number, $N_t$.~{We observe that even for small
attractions and for relatively small atom numbers of $N_t\lesssim350$, $498$, and $279$, in
Figs.~\ref{fig:chem_pot}(a)-(c), respectively, $\mu_{t}<0$, thus verifying the bound state
nature of the entire system}.~However, its behavior for increasing $N_t$ depends strongly on
both the interactions and the particle imbalance.

Particularly, the chemical potential of the minority component (i.e., $\mu_2$) acquires
negative values which become larger in magnitude as $N_t$ increases {(see,
Figs.~\ref{fig:chem_pot}(a)-(b) and (d)-(f)) with the exception of simultaneous weak
attractions and low particle imbalance, as is shown in Fig.~\ref{fig:chem_pot}(c)}.~This
trend unveils the bound character of the minority component and thus its droplet nature which
is more prominent for either fixed particle imbalance and larger intercomponent attraction,
e.g., compare Figs.~\ref{fig:chem_pot}(a) and (d), or fixed interactions and larger imbalance,
see for instance Fig.~\ref{fig:chem_pot}(d) and (f).~On the other hand, the chemical potential
of the majority component, $\mu_1$, is positive, and shows a gradual increase for larger $N_t$,
a behavior that is independent of the interactions or the atom number.~This is attributed to
the existence of the excess particles in the majority component becoming more prominent as $N_t$
increases (see, also Fig.~\ref{fig:den_profs}), and essentially acting against the bound state
nature of the ensuing configurations.~Hence, the interplay of the chemical potentials of the two
components may lead to either a many-body bound state for strong intercomponent attractions as
can be seen in Figs.~\ref{fig:chem_pot}(d)-(f), or to a transition of the entire system from a
droplet to a trapped gas phase~\cite{Flynn_trap_drops} at weaker attractions as $N_t$ increases,
see Figs.~\ref{fig:chem_pot}(a)-(c).~Therefore, it is possible to control the bound state character
of the many-body system.

It becomes also evident that interactions play a crucial role in the aforementioned
transition, i.e., from negative to positive chemical potential.~This crossing is
shifted to significantly larger atom numbers for stronger intercomponent attractions,
see, e.g., Figs.~\ref{fig:chem_pot}(b) and (e).~As such, the many-body state cannot
maintain its self-bound droplet character throughout the $N_t$ variation.~Additionally,
an enhanced intercomponent imbalance such as $N_1=80\%N_t$, $N_2=20\%N_t$ sustains the
bound character for a wider particle number interval as can be deduced by inspecting
Figs.~\ref{fig:chem_pot}(a) and (c).~For example, the crossing to positive values occurs
at $N_{t}\approx 351$ in Fig.~\ref{fig:chem_pot}(a) [$N_{t}\approx 280$ in Fig.~\ref{fig:chem_pot}(c)]
for a $N_1=80\%N_t$, $N_2=20\%N_t$ [$N_1=60\%N_t$, $N_2=40\%N_t$] imbalance.~This can be
traced back to the fact that a smaller imbalance is associated with larger total chemical
potential, and thus energy of the system, e.g., compare Figs.~\ref{fig:chem_pot}(a) and (c)
or Figs.~\ref{fig:den_profs}(d) and (f).~This indeed confirms that strongly particle imbalanced
droplets correspond to energetically lower lying configurations.
{~Also, it is worth mentioning
that considering fixed particle imbalance while tuning the involved interactions such that the mean-field balance point $\delta g$ increases accelerates the transition to the gas phase, compare for instance,
Figs.~\ref{fig:chem_pot}(a), (b) and (d) having $\delta g \approx 0.9$, 0.6 and 0.2 respectively.~However, keeping the particle imbalance and $\delta g$ constant but reducing one of the intracomponent repulsions and the intercomponent attraction leads faster to the gas state, see Fig.~\ref{fig:chem_pot}(d), (f).}~Summarizing, the droplet phase is parametrically extended
either for fixed particle imbalance and stronger attractions or constant interactions and smaller
imbalance, see Fig.~\ref{fig:chem_pot}.~We also remark that this transition occurs at smaller
$N_t$ for increasing trap strength (results not shown) since, in general, the trap does not favor
the bound state formation.~This was also demonstrated for the respective single-component droplets
of a symmetric mixture~\cite{Mistakidis_formation}.

\subsection{Excitation spectrum of particle imbalanced droplets}\label{sec:BdG}

\begin{figure}[!pt]
\begin{center}
\begin{overpic}[trim={1.5cm 0 0.5cm 0},height=.151\textheight, angle =0]{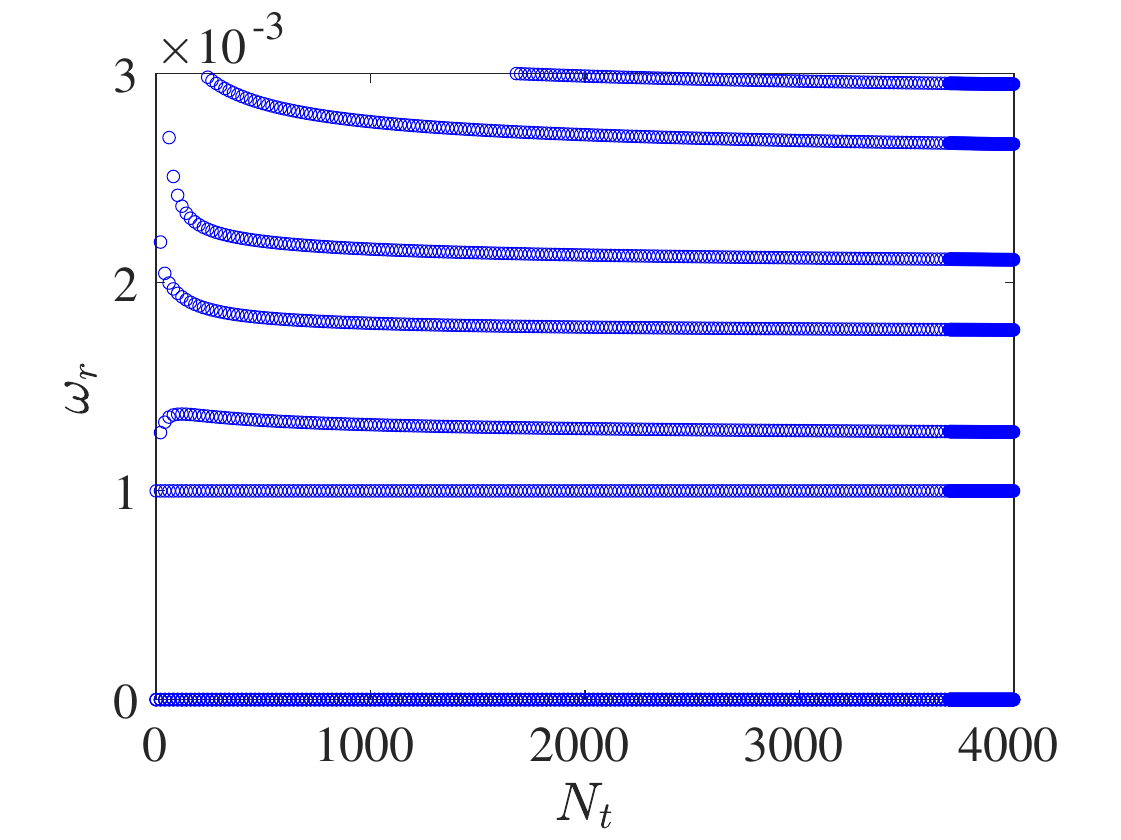}
\put(9,20){{\small $\textrm{(a)}$}}
\end{overpic}
\begin{overpic}[trim={1cm 0 1cm 0},height=.151\textheight, angle =0]{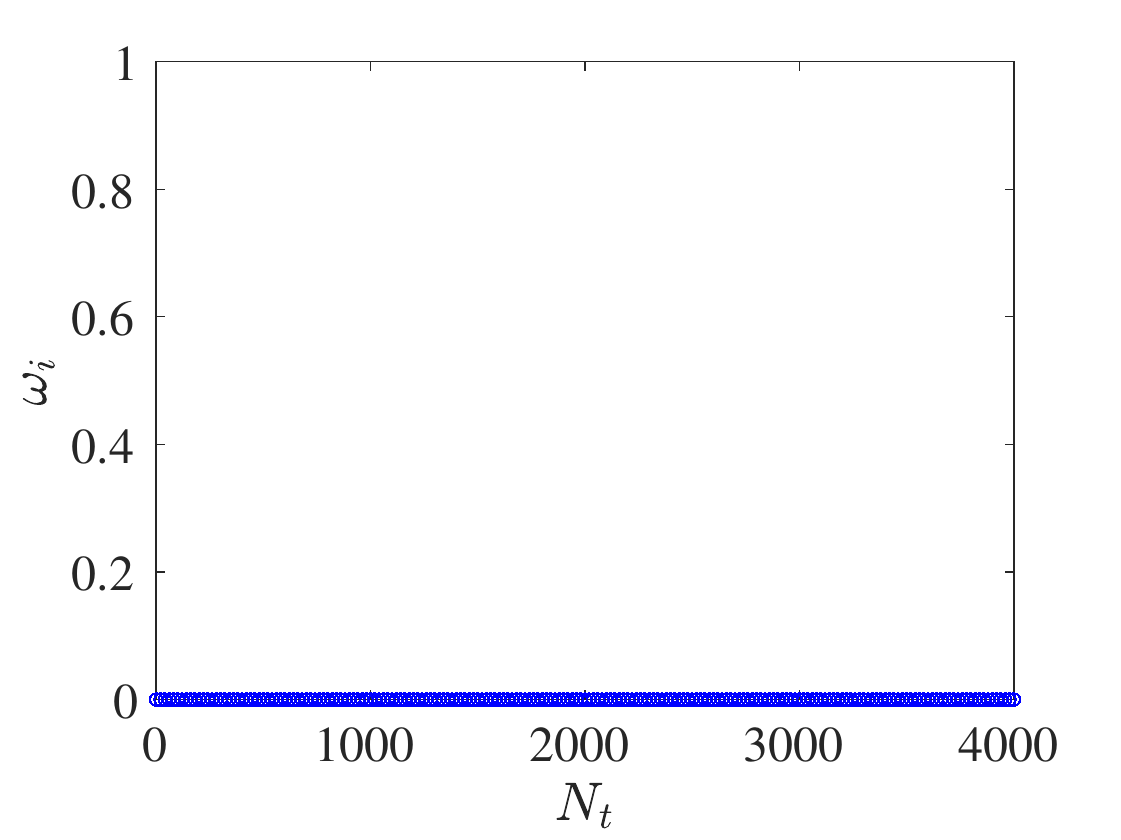}
\put(12,70){{\small $\textrm{(b)}$}}
\end{overpic}\\
\begin{overpic}[trim={1.5cm 0 0.5cm 0},height=.151\textheight, angle =0]{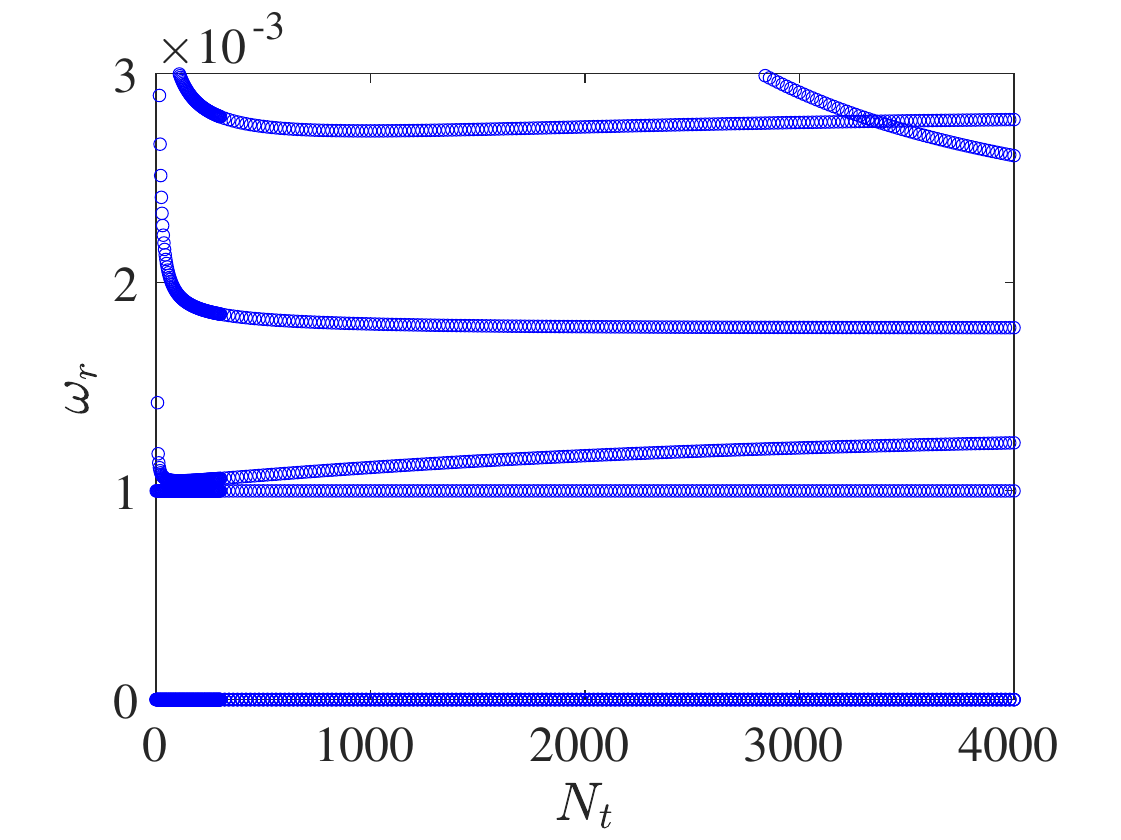}
\put(9,20){{\small $\textrm{(c)}$}}
\end{overpic}
\begin{overpic}[trim={1cm 0 1cm 0},height=.151\textheight, angle =0]{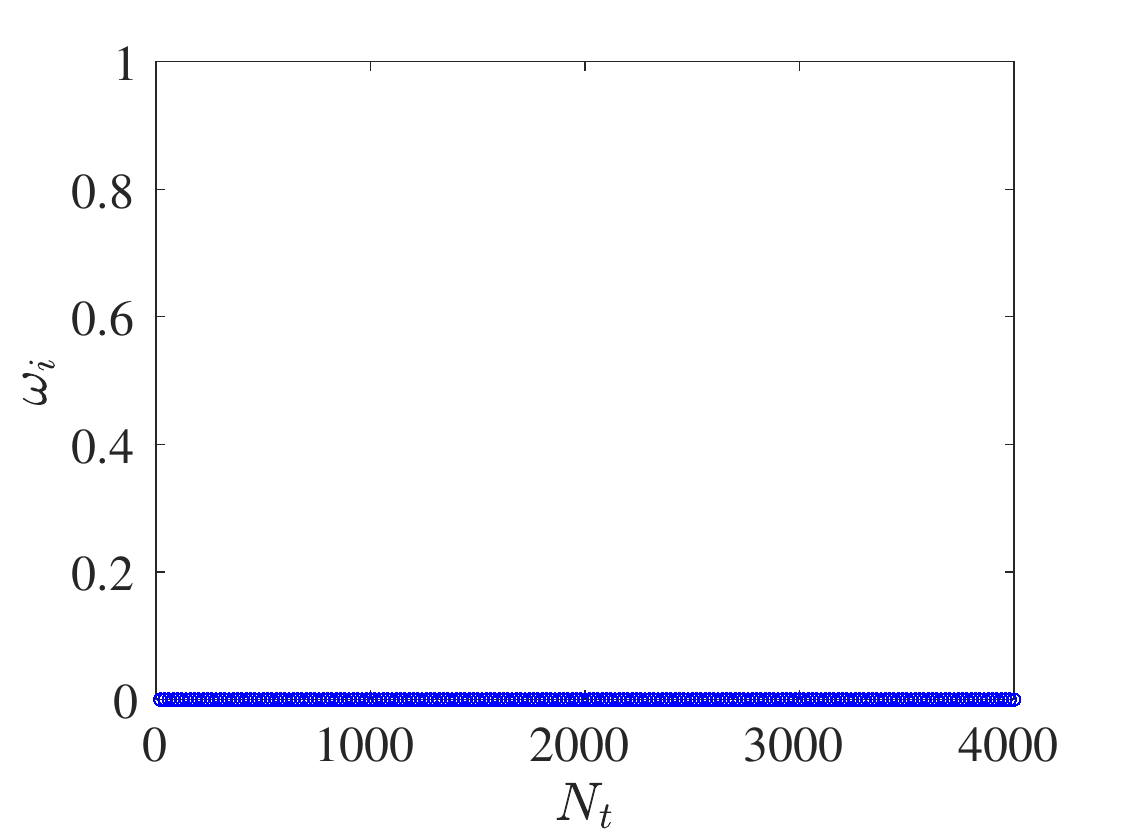}
\put(12,70){{\small $\textrm{(d)}$}}
\end{overpic}
\end{center}
\caption{
\label{fig:Bdg_8020}
(Color online)
(a), (c) Real and (b), (d) imaginary parts of the BdG excitation spectra for particle
imbalanced ($N_1=80\%N_t$, $N_2=20\%N_t$), droplet configurations upon varying the total
atom number $N_t$.~The scale of the real spectrum being of the order of $10^{-3}$ reflects
the trap frequency $\Omega=10^{-3}$.~The two-component settings feature intracomponent
repulsion $g_{1}=g_{2}=1$, and intercomponent attraction (a), (b) $g_{12}=-0.1$ and (c),
(d) $g_{12}=-0.8$.~Stability of the stationary two-component droplet configurations can be
inferred due to the vanishing imaginary part ($\omega_i=0$) of the spectrum, irrespectively
of the atom number and the interactions. All parameters are dimensionless.}
\end{figure}

To infer the spectral stability of the above-described two-component droplet states, we next
rely on the so-called BdG  analysis~\cite{Bogolyubov-1947}.~We remark that the excitation
spectrum of only 1D symmetric droplet settings has been examined thus far~\cite{Tylutki,katsimiga2023interactions} appreciating also the effect of nonlinear excitations~\cite{Katsimiga_sol_drops} and spin-orbit coupling~\cite{Gangwar_2024}. However, the excitation spectrum of genuine two-component droplet
setups that we aim to investigate herein remains highly unexplored.

Specifically, we perturb the obtained stationary solutions, $\psi^{(0)}_{j}(x)$
by employing the ansatz
\begin{align}
\widetilde{\psi}_{j}=e^{-\ii\mu_{j} t}%
\left[\psi^{(0)}_{j}+\varepsilon\left(a_{j}e^{\ii\omega t}+b_{j}^{\ast}e^{-\ii\omega^{\ast}t}\right)\right],%
~\varepsilon \ll 1.
\label{pertr_ansatz_2C}
\end{align}
{Here, $a_{j}(x)$ and $b_{j}(x)$ form the eigenvector $[a_{j}(x),b_{j}(x)]^{T}$
associated with the eigenfrequency $\omega$.~Upon inserting Eq.~\eqref{pertr_ansatz_2C}
into Eq.~\eqref{eq:eGPE}, we arrive at order $\mathcal{O}(\varepsilon)$ at the operator
eigenvalue problem given by Eq.~\eqref{eq:stab} in Appendix~\ref{sec:Appendix1}.~Its
solution yields the eigenfrequencies $\omega=\omega_{r}+\ii\,\omega_{i}$ through which the
stability of the pertinent two-component states can be deduced.~Recall that a solution is
classified as spectrally stable if all the eigenfrequencies $\omega$ lie on the real axis,
i.e., there are no eigenfrequencies with a non-zero imaginary part. On the contrary, if there
exist eigenfrequencies with a positive imaginary part, i.e., $\omega_{i}>0$, then the solution
is deemed unstable. It should be noted that the eigenfrequencies $\omega$ connect with
the eigenvalues $\lambda$ through a rotation of the spectral plane by $\pi/2$, or, $\omega=\ii \lambda$}.

\begin{figure}[!pt]
\begin{center}
\begin{overpic}[trim={1.5cm 0 0.5cm 0},height=.151\textheight, angle =0]{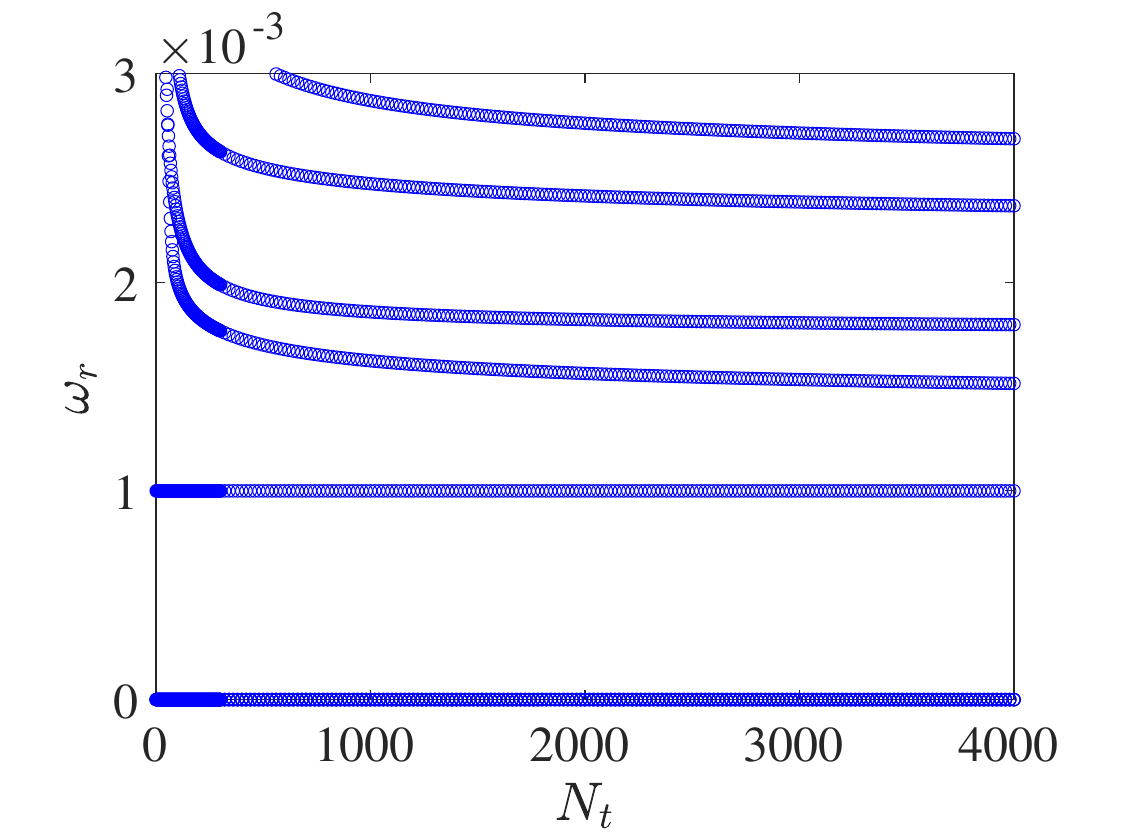}
\put(9,20){{\small $\textrm{(a)}$}}
\end{overpic}
\begin{overpic}[trim={1cm 0 1cm 0},height=.151\textheight, angle =0]{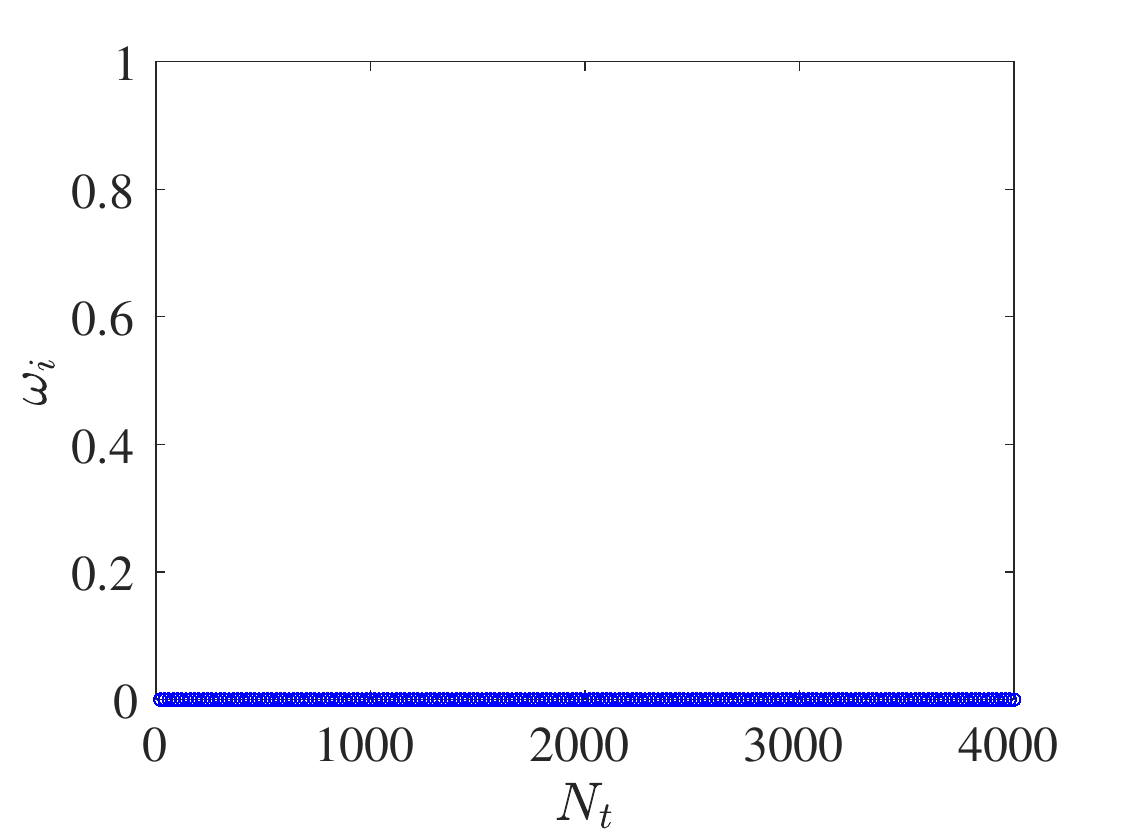}
\put(12,70){{\small $\textrm{(b)}$}}
\end{overpic}\\
\begin{overpic}[trim={1.5cm 0 0.5cm 0},height=.151\textheight, angle =0]{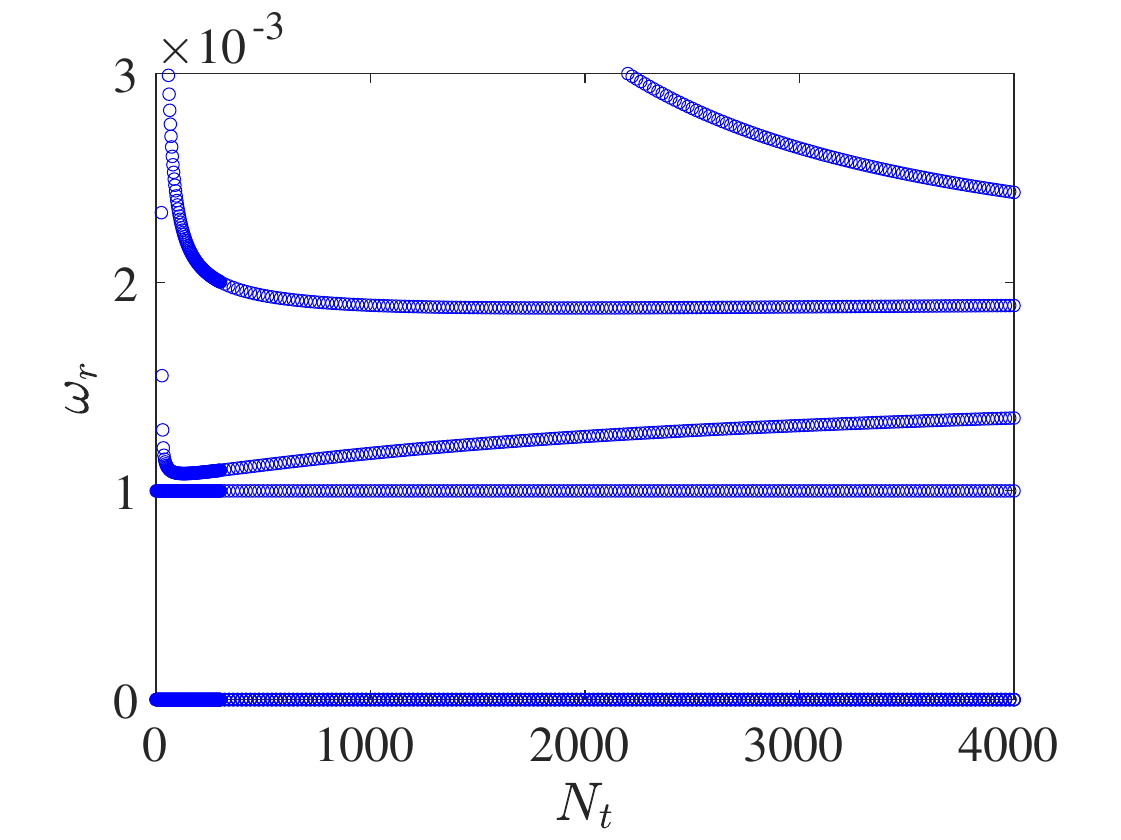}
\put(9,20){{\small $\textrm{(c)}$}}
\end{overpic}
\begin{overpic}[trim={1cm 0 1cm 0},height=.151\textheight, angle =0]{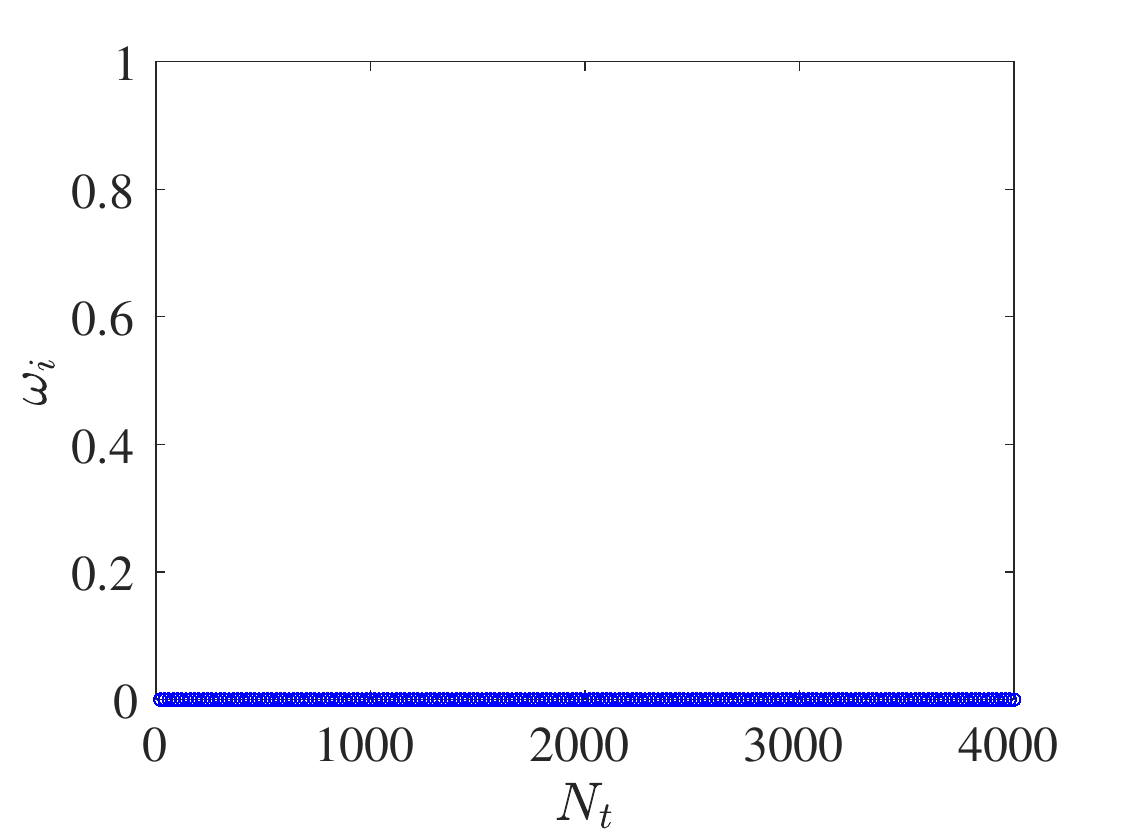}
\put(12,70){{\small $\textrm{(d)}$}}
\end{overpic}
\end{center}
\caption{
\label{fig_Bdg_6040}
(Color online)
Excitation spectrum of a two-component bosonic droplet system with particle imbalance
($N_1=60\%N_t$, $N_2=40\%N_t$) with respect to $N_t$.~Both the (a), (c) real and (b),
(d) imaginary parts of $\omega$ are depicted with $g_{1}=g_{2}=1$, while (a), (b)
$g_{12}=-0.1$ and (c), (d) $g_{12}=-0.8$.~Spectral stability of the ensuing two-component
droplet states is identified by the absence of imaginary parts in $\omega$, i.e.,
$\omega_i=0$, for varying atom number and distinct interactions.~All quantities are
dimensionless.}
\end{figure}

\begin{figure*}[!pt]
\begin{center}
\begin{overpic}[trim={0.5cm 0 0.5cm 0},height=.15\textheight, angle =0]{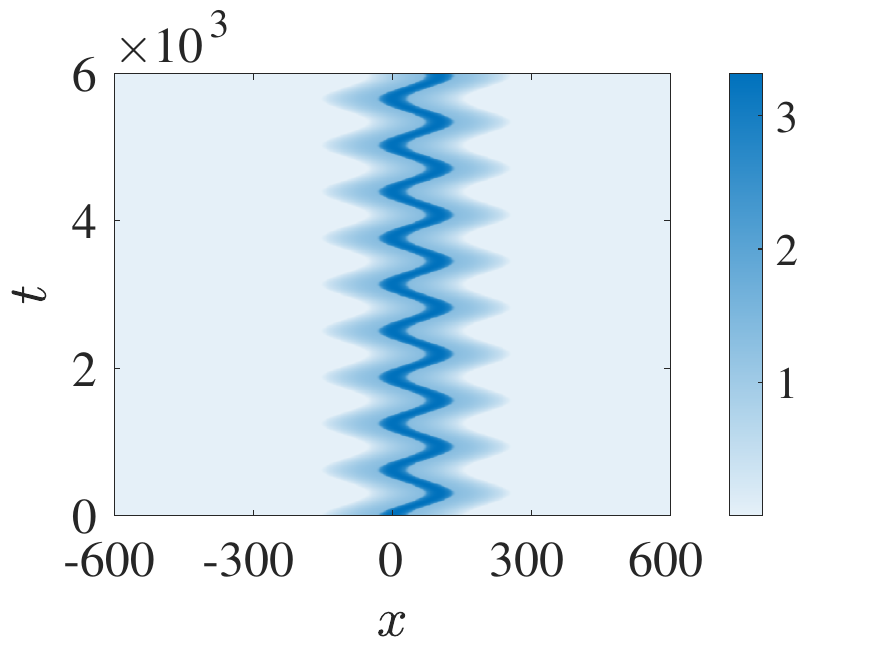}
\put(11,64){{\small $\textrm{(a)}$}}
\end{overpic}
\begin{overpic}[trim={0.5cm 0 0.5cm 0},height=.15\textheight, angle =0]{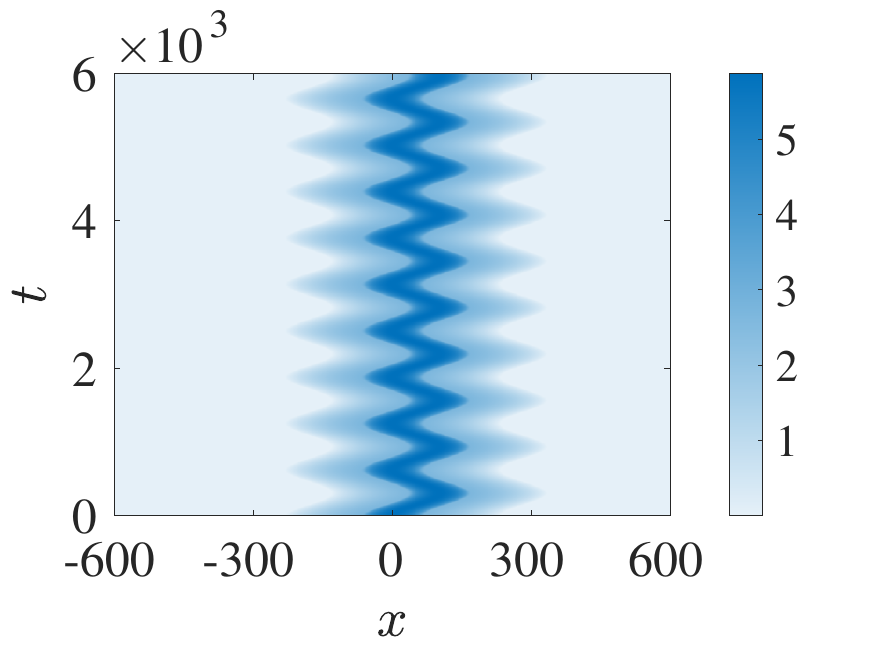}
\put(11,64){{\small $\textrm{(b)}$}}
\end{overpic}
\begin{overpic}[trim={0.5cm 0 0.5cm 0},height=.15\textheight, angle =0]{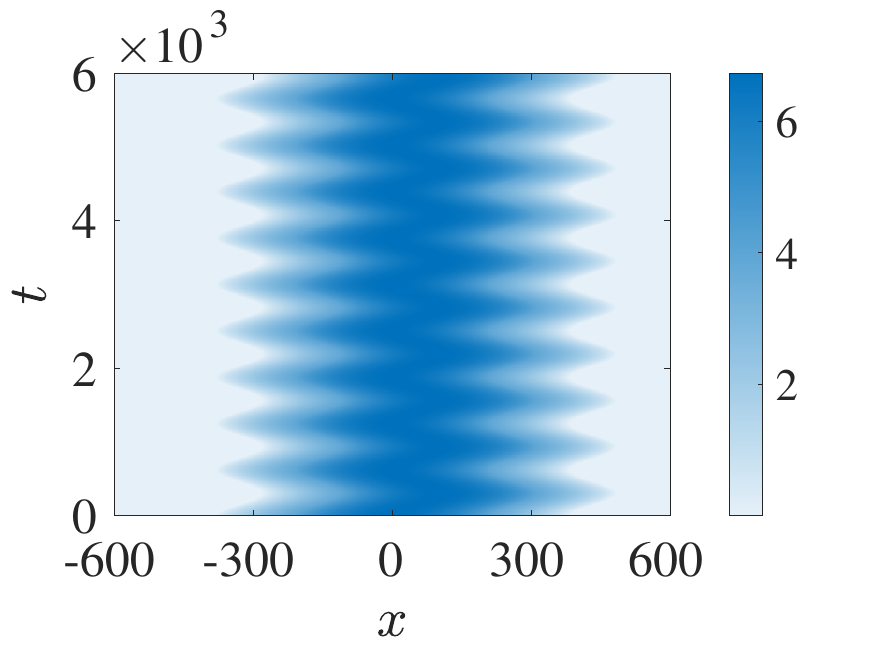}
\put(11,64){{\small $\textrm{(c)}$}}
\end{overpic}
\begin{overpic}[trim={0.5cm 0 0.5cm 0},height=.15\textheight, angle =0]{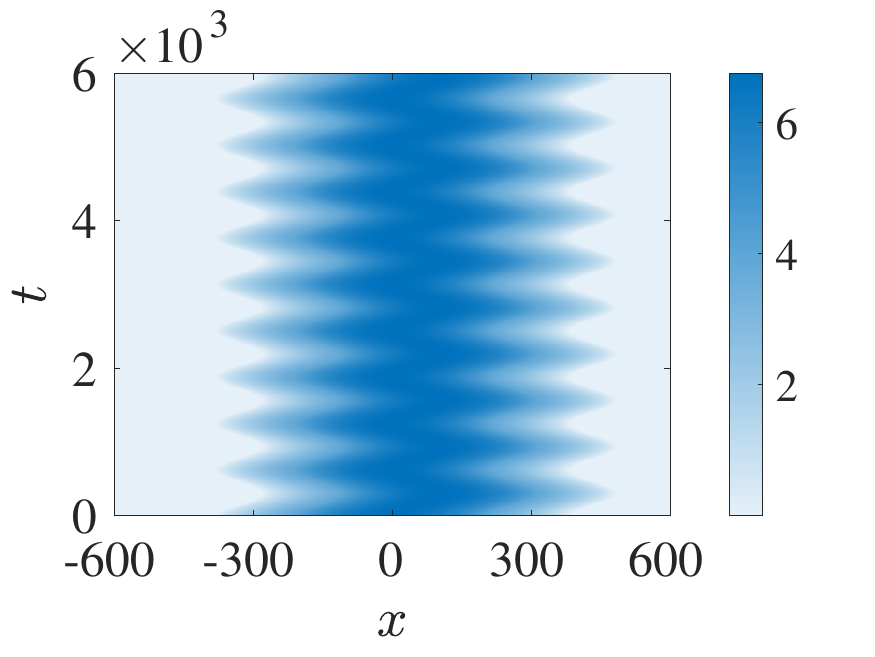}
\put(11,64){{\small $\textrm{(d)}$}}
\end{overpic}\\
\begin{overpic}[trim={0.5cm 0 0.5cm 0},height=.15\textheight, angle =0]{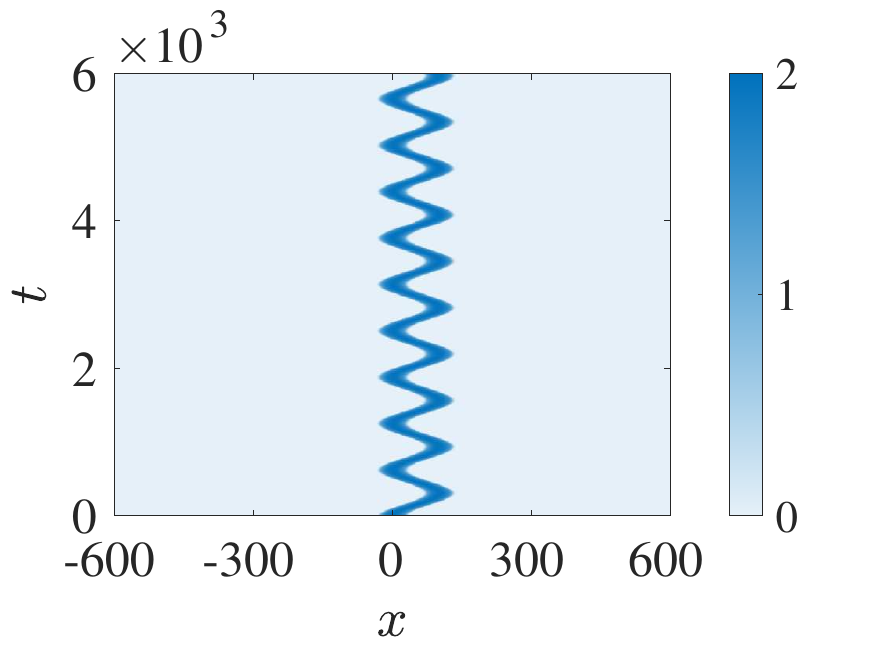}
\put(11,64){{\small $\textrm{(e)}$}}
\end{overpic}
\begin{overpic}[trim={0.5cm 0 0.5cm 0},height=.15\textheight, angle =0]{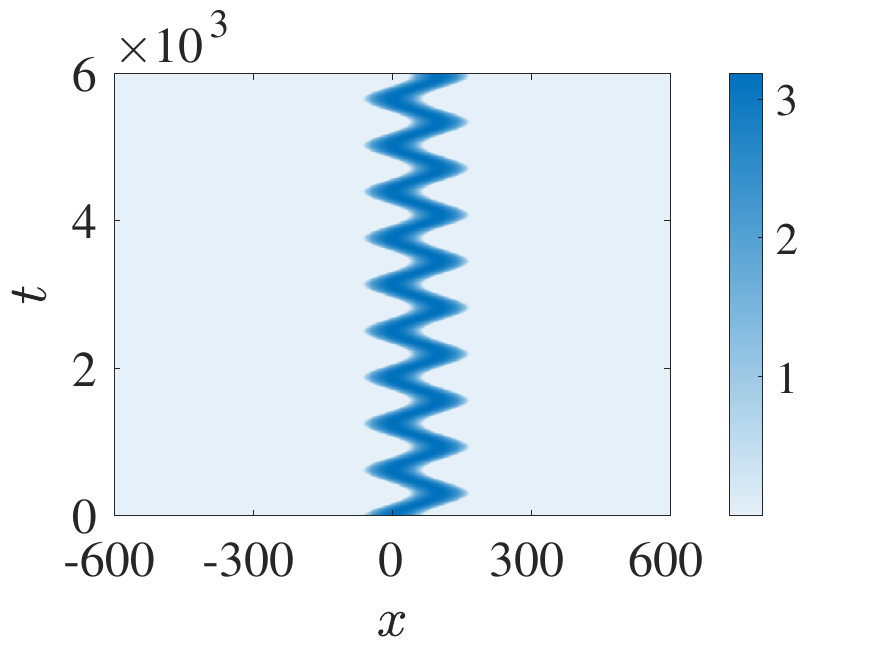}
\put(11,64){{\small $\textrm{(f)}$}}
\end{overpic}
\begin{overpic}[trim={0.5cm 0 0.5cm 0},height=.15\textheight, angle =0]{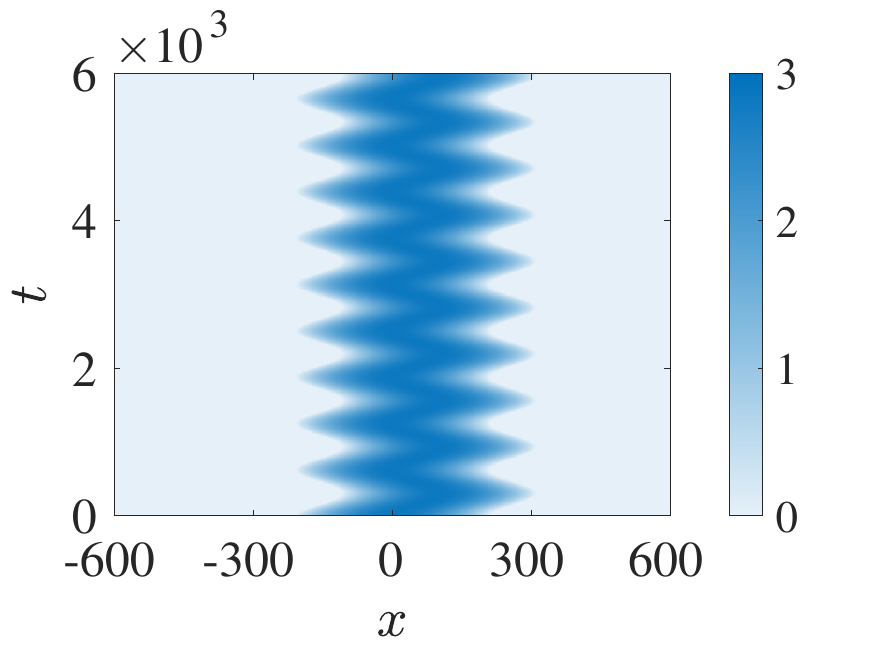}
\put(11,64){{\small $\textrm{(g)}$}}
\end{overpic}
\begin{overpic}[trim={0.5cm 0 0.5cm 0},height=.15\textheight, angle =0]{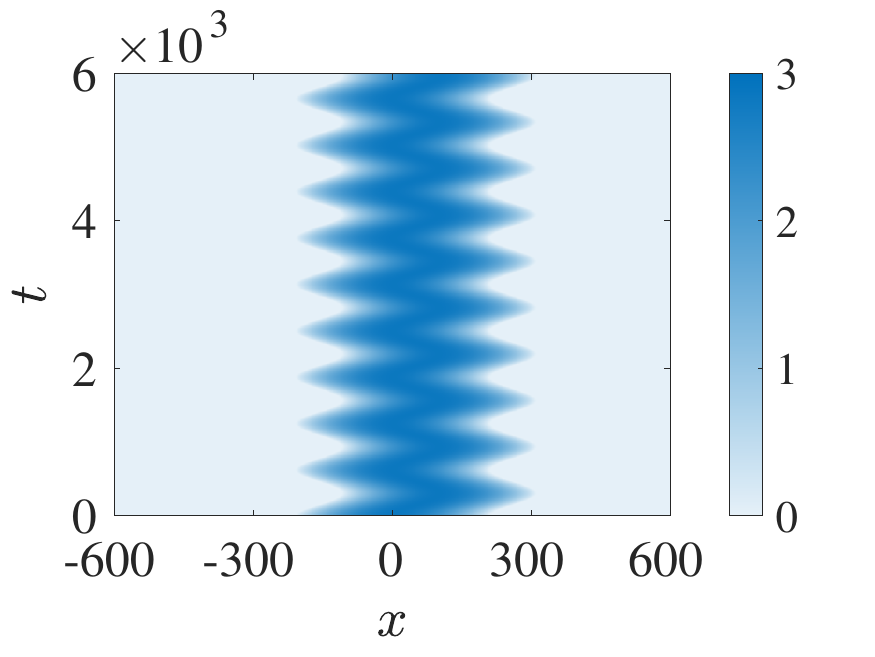}
\put(11,64){{\small $\textrm{(h)}$}}
\end{overpic}
\end{center}
\caption{
\label{fig:dipole}
(Color online)
Time evolution of the droplet densities of each component following a sudden change
of the trap's center from $x_0=0$ to $x_0=+50$.~The intercomponent particle imbalance
is fixed to ($N_1=80\%N_t$, $N_2=20\%N_t$), while the total atom number corresponds
to (a), (e) $N_t=500$, (b), (f) $N_t=1500$, (c), (g) $N_t=3980$ and (d), (h) $N_t=3999$.%
~Apparently, both the majority (upper panels) and the minority (lower panels) components
perform a collective dipole motion reflected by their in-phase oscillation within the trap.%
~In particular, for lower atom numbers where the minority is highly localized imprints
its motion to the majority component cloud.~In all cases, the two-component bosonic droplet
is initiated in its ground state configuration with interactions $g_{1}=g_2=1$ and $g_{12}=-0.8$,
while being under the influence of an external harmonic trap with frequency $\Omega=10^{-2}$.%
~All quantities shown are in dimensionless units.}
\end{figure*}

The resulting discrete excitation spectra of the highly- and weakly-imbalanced systems referring
to ($N_1=80\%N_t$, $N_2=20\%N_t$) and ($N_1=60\%N_t$, $N_2=40\%N_t$) respectively are illustrated in
Fig.~\ref{fig:Bdg_8020} and Fig.~\ref{fig_Bdg_6040} as a function of $N_{t}$, and for different
intercomponent attractions.~Note that we obtain the respective BdG spectra by numerically solving
the eigenvalue problem of Eq.~\eqref{eq:stab} for the stationary solutions presented in Fig.~\ref{fig1_summary}.%
~Specifically, Figs.~\ref{fig:Bdg_8020} and~\ref{fig_Bdg_6040} depict the dependence
of the real and imaginary parts of the 
eigenfrequencies, i.e., $\omega_{r}$ and
$\omega_{i}$, respectively, on $N_{t}$.~It can be readily deduced that all two-component droplets
are spectrally stable, a finding that is supported by the absence of a (positive) imaginary part,
i.e., $\omega_i=0$ (being of the order of $\approx  10^{-7}$ here) in the BdG spectrum, see, panels
(b) and (d) of Figs.~\ref{fig:Bdg_8020} and~\ref{fig_Bdg_6040}.~We remark that we have independently
verified the validity of our BdG stability results by performing direct dynamical evolution of the
above-discussed droplet solutions.~The configurations remain intact for times up to $t=6000$ that
we have checked.

The real part of the BdG spectrum entails information regarding the collective modes of the
droplets such as the dipole and the breathing modes appearing at $\omega_r=\Omega$ and
$\omega_r \approx \sqrt{3}\Omega$ for $N_{t}\gg 1$, respectively.~These can be traced by monitoring the
energetically lowest lying trajectory, and the one around $\sqrt{3}\Omega$ shown in panels
(a) and (c) of Figs.~\ref{fig:Bdg_8020} and~\ref{fig_Bdg_6040}.~Here, it is worth pointing out
that the values of the collective modes (except of the dipole) depend on $N_t$, and hence on the
interactions, such as the one of the breathing mode whose thermodynamically predicted
value~\cite{abraham2014quantum,Englezos_trap} is seen here to be reached for $N_t>3000$.~The
consistency of our BdG calculations is further supported by the fact that the dipolar mode of
frequency  $\omega_{r}=\Omega(=10^{-3})$ is persistent in our spectra.~From the symmetries point
of view, a careful inspection of the spectra reveals the existence of $4$ zero, or equivalently,
of $2$ pairs of zero eigenfrequencies associated with the phase invariance (i.e., $U(1)$ symmetry)
of each of the two components.~Recall that the space translation invariance is broken due to
the presence of the external potential $V(x)$, i.e., the eGPE system is no longer translationally
invariant in $x$.

\begin{figure*}[!pt]
\begin{center}
\begin{overpic}[trim={0.5cm 0 0.5cm 0},height=.15\textheight, angle =0]{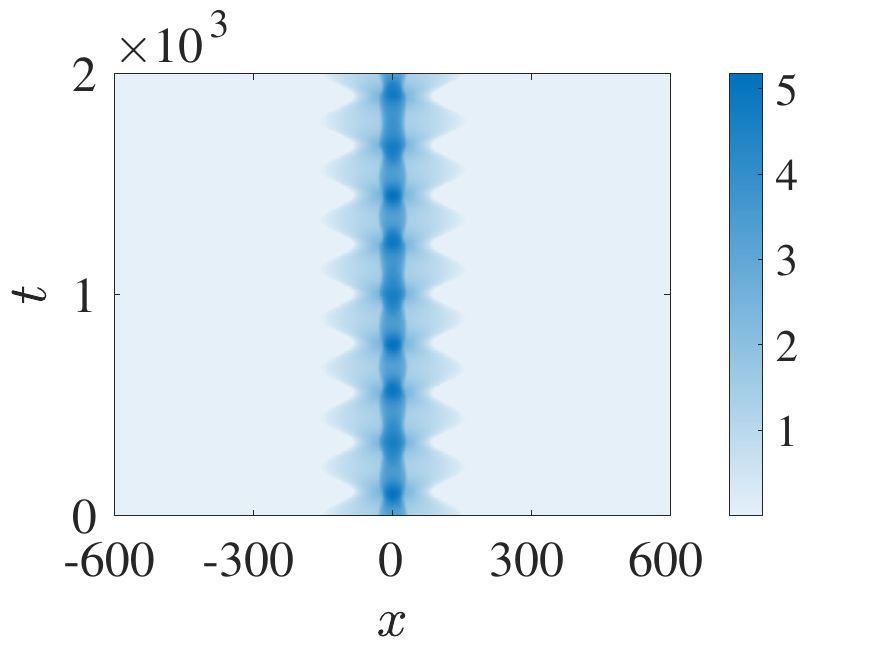}
\put(11,64){{\small $\textrm{(a)}$}}
\end{overpic}
\begin{overpic}[trim={0.5cm 0 0.5cm 0},height=.15\textheight, angle =0]{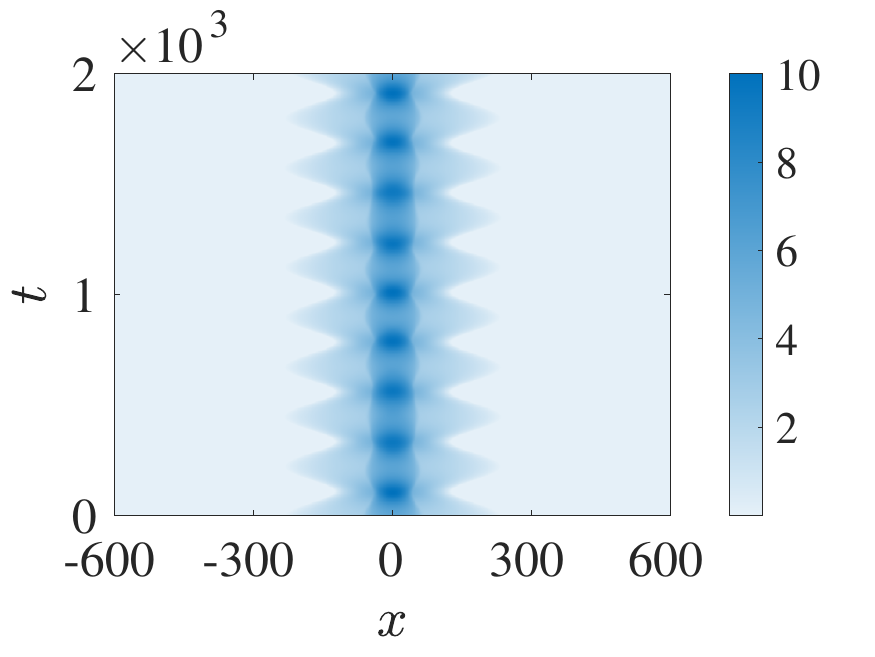}
\put(11,64){{\small $\textrm{(b)}$}}
\end{overpic}
\begin{overpic}[trim={0.5cm 0 0.5cm 0},height=.15\textheight, angle =0]{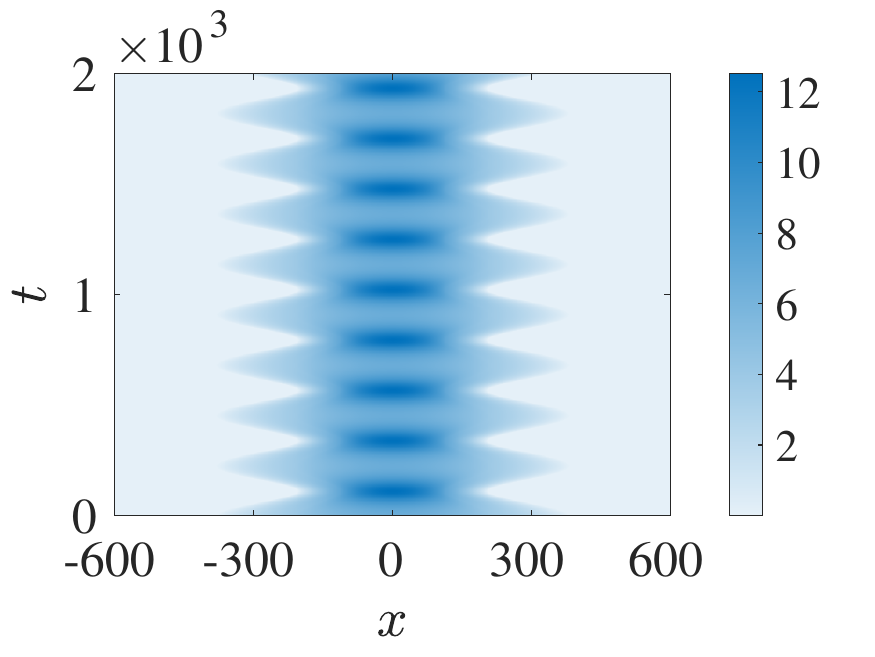}
\put(11,64){{\small $\textrm{(c)}$}}
\end{overpic}
\begin{overpic}[trim={0.5cm 0 0.5cm 0},height=.15\textheight, angle =0]{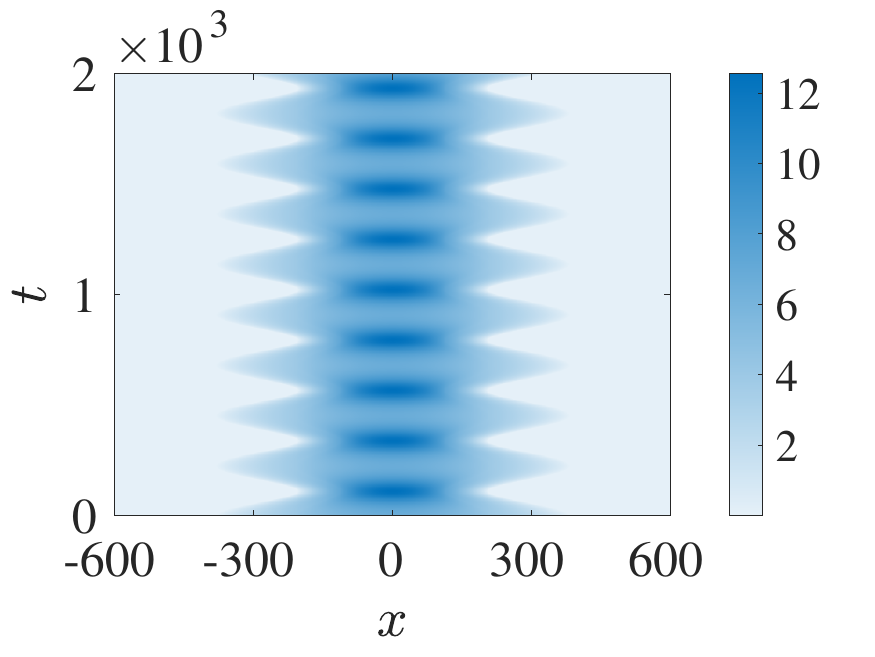}
\put(11,64){{\small $\textrm{(d)}$}}
\end{overpic}\\
\begin{overpic}[trim={0.5cm 0 0.5cm 0},height=.15\textheight, angle =0]{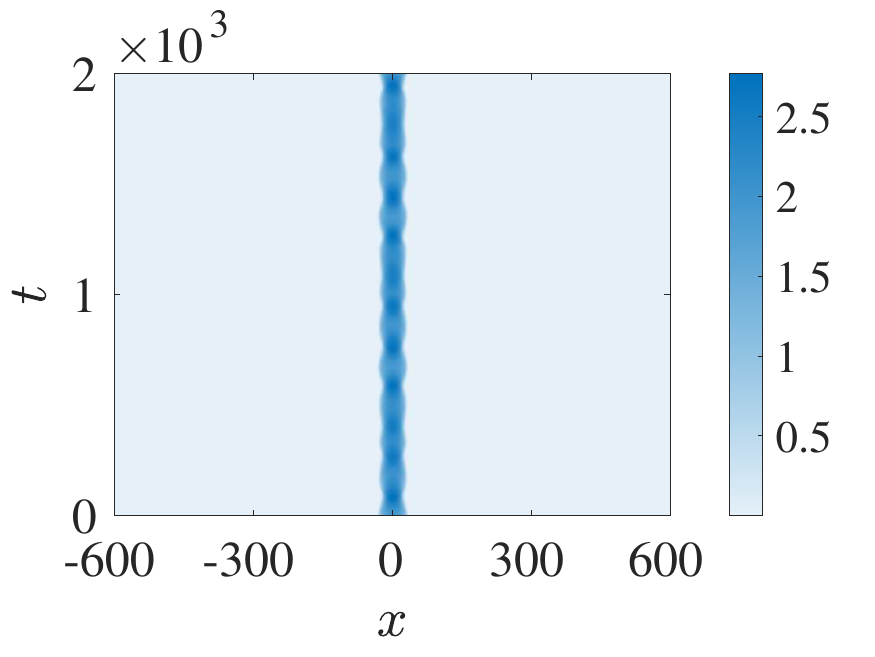}
\put(11,64){{\small $\textrm{(e)}$}}
\end{overpic}
\begin{overpic}[trim={0.5cm 0 0.5cm 0},height=.15\textheight, angle =0]{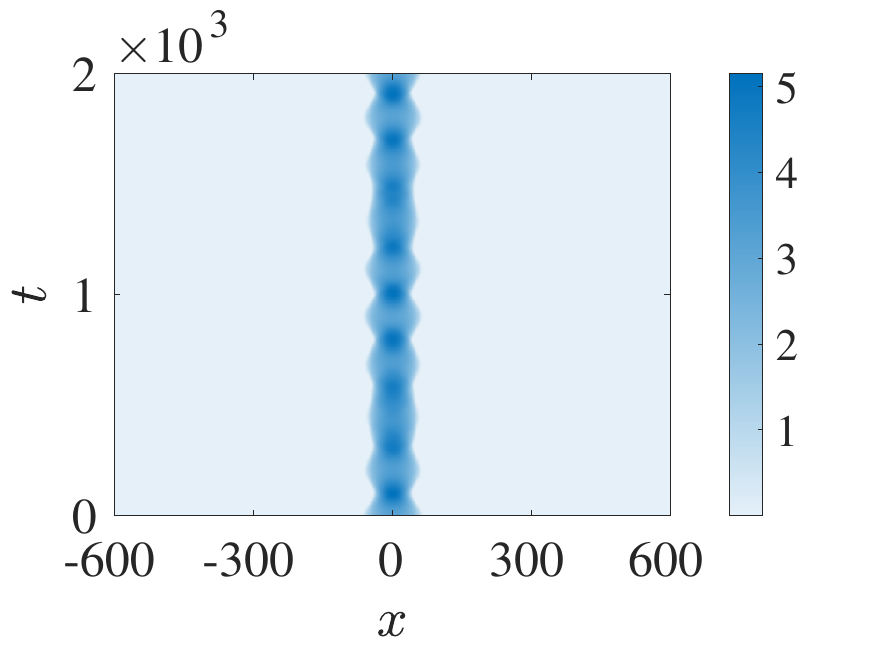}
\put(11,64){{\small $\textrm{(f)}$}}
\end{overpic}
\begin{overpic}[trim={0.5cm 0 0.5cm 0},height=.15\textheight, angle =0]{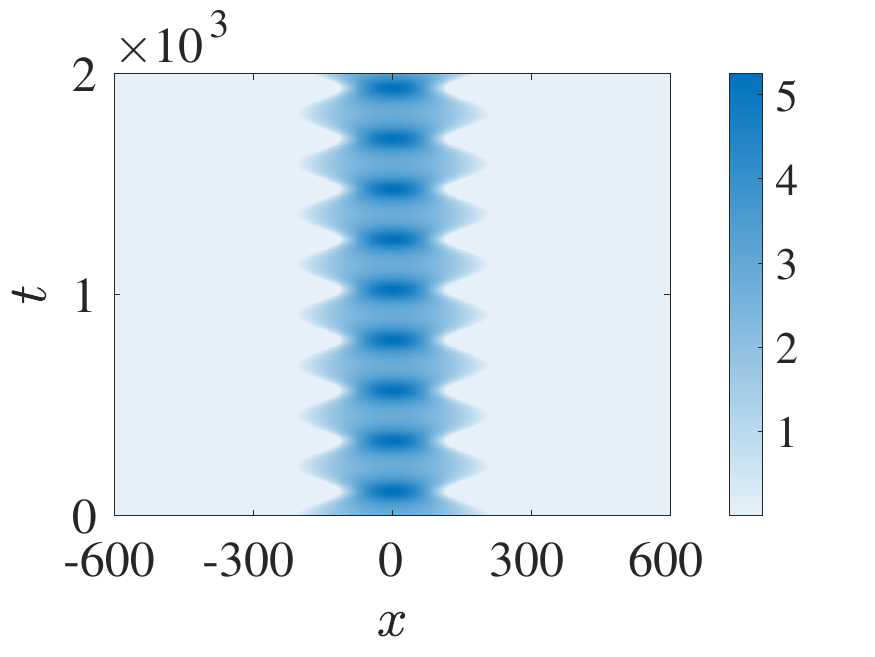}
\put(11,64){{\small $\textrm{(g)}$}}
\end{overpic}
\begin{overpic}[trim={0.5cm 0 0.5cm 0},height=.15\textheight, angle =0]{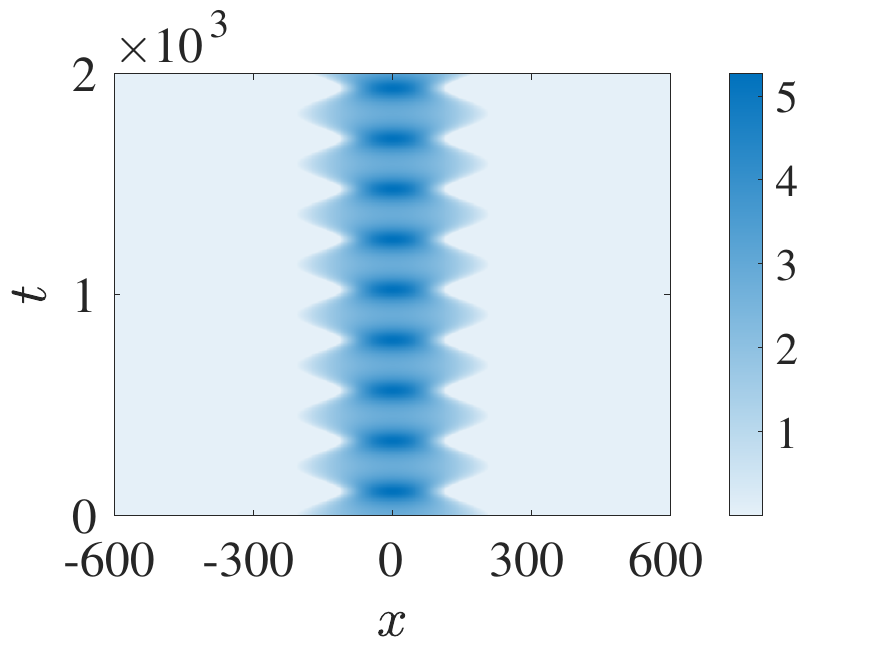}
\put(11,64){{\small $\textrm{(h)}$}}
\end{overpic}
\end{center}
\caption{
\label{fig:breathing}
(Color online) Dynamics of the individual component droplet density distributions
after a quench of the trap frequency from $\Omega=10^{-2}$ to $\Omega_f=\pi \Omega/2 \approx 1.6 \times 10^{-2}$.%
~Different configurations are depicted associated with (a), (e) $N_t=500$, (b),
(f) $N_t=1500$, (c), (g) $N_t=3980$ and (d), (h) $N_t=3999$.~The quench triggers a
periodic expansion and contraction dynamics of both the majority (upper panels) and
the minority (lower panels) clouds.~Both the droplet and excess atom segments of the
majority component move in-sync and the back action of the highly localized  minority
component to the majority one is evident for smaller atom numbers.~The system is prepared
in its ground state characterized by ($N_1=80\%N_t$, $N_2=20\%N_t$), $g_{1}=g_2=1$,
$g_{12}=-0.8$ and $\Omega=10^{-2}$. Quantities are presented in dimensionless units.}
\end{figure*}

\section{Collective dynamics of two-component droplets}\label{sec:dynamics}

Having analyzed the stationary and stability properties of the particle imbalanced
two-component droplet configurations, we subsequently turn our attention to their
dynamical behavior.~Our aim, here, is in part to verify the predictions of the spectrum
{associated with specific modes,}
and also visualize basic characteristics of the droplets response that will pave the
path for future investigations.~As such, the energetically lower fundamental collective
modes of the system are studied referring to the droplets dipole and breathing motion.~These
modes are triggered through appropriate quenches of the external trap, ensuring also that
the emergent dynamics can be adequately described by the coupled set of eGPEs.~This is because
higher-order correlations do not play a significant role as it was argued for symmetric
droplets~\cite{Mistakidis_formation,Englezos_trap}.%
~Namely, a sudden shift of the trap's center results in exciting the dipole motion, while
an abrupt change of the trap's frequency induces the breathing dynamics of the  components.

Below, we present the time-evolution of representative cases of two-component droplets
corresponding to different total atom numbers while maintaining the imbalance ($N_1=80\%N_t$, $N_2=20\%N_t$)
and interactions ($g_{1}=g_{2}=1$, $g_{12}=-0.8$), see also Fig.~\ref{fig1_summary} and
Fig.~\ref{fig:den_profs}.~A similar phenomenology occurs for other intercomponent interactions
(not shown for brevity), while as the particle imbalance is decreased and the system tends to
acquire equal intercomponent population, we retrieve the known characteristics of the symmetric
droplet mixture reported, for instance, in Refs~\cite{Parisi_QMC,Englezos_trap}.~To better visualize
the emergent collective dynamics in somewhat smaller timescales, we use a slightly tighter trap,
as compared to the above sections, characterized by $\Omega=10^{-2}$.

The spatiotemporal density dynamics of both components after a quench of each component's trap
center from $x_0=0$ to $x_0=+50$ for $N_t=500$, $N_t=1500$, $N_t=3980$ and $N_t=3999$ is depicted
in Fig.~\ref{fig:dipole}.~In all cases, both components perform a collective in-trap  oscillatory
motion with frequency equal to the one of the trap.~This is a clear manifestation of the excitation
of the dipole mode.~Notice that the dipole mode frequency is measured through the spectrum of the
time-evolved center-of-mass of each individual cloud.~It is found that independently of $N_t$ or
the interactions, and hence the ensuing droplet configurations, the dipole mode frequency is unchanged
which further confirms the BdG outcome of Figs.~\ref{fig:Bdg_8020}(c) and (d), where $\omega_r=\Omega$.%
~Interestingly, for smaller $N_t$, where the spatial localization of the minority component distribution
is more enhanced and the droplet fraction smaller, we observe a prominent back action to the majority
component as a consequence of the intercomponent attraction.~Here, a density hump builds upon the majority
component distribution imprinting the minority's  trajectory.~In this case, both the droplet and excess
atom fragments of the majority component oscillate in-phase, see Figs.~\ref{fig:dipole}(a), (b), (e),
and (f).~This suggests that the droplet fragment of the composite configuration drifts the gaseous one.~On
the other hand, and for larger $N_t$ (where the droplet fraction increases), it can be seen that the
aforementioned two segments of the majority component are hardly discernible, and the cloud undergoes an
overall oscillatory behavior as is shown in Figs.~\ref{fig:dipole}(c), (d), (g), and (h).

The breathing mode of the droplet is triggered by quenching the trap's frequency, and corresponds
to a collective expansion and contraction of each component distribution in a periodic manner.%
~The time-evolution of the densities of both participating components are illustrated in
Figs.~\ref{fig:breathing} for the same atom numbers used for the dipole mode.~We estimate
the underlying breathing mode frequency by monitoring the dynamics of the width of the individual
wavepackets~\cite{Ronzheimer,fukuhara2013quantum}.
~We find that the
breathing mode frequency exhibits an interaction dependent behavior for both components, and in
particular, it features a slightly decreasing tendency for larger $N_t$ while being eventually
saturated to $\approx\sqrt{3}\Omega$ for $N_t>3000$.~Interestingly, the minority component for
relatively small $N_t$, see Figs.~\ref{fig:breathing}(e) and (f), has two distinguishable breathing
frequencies one of them coinciding with the breathing frequency of the majority component and the
other being slightly larger.~This two-mode response of the minority hints towards a few-body effect
which becomes suppressed for larger $N_t$, see Figs.~\ref{fig:breathing}(g), (h).

This overall behavior but also the values of the breathing frequency  extracted from the dynamics match
the predictions of the BdG spectrum. Furthermore, as in the case of the dipole mode dynamics discussed
above, the back action of the spatially localized minority component into the majority one is well discernible
for relative small $N_t$.~This can be readily seen in the density evolution of the majority component
presented in Figs.~\ref{fig:breathing}(a), (b), (e), and (f) where traces of the minority distribution
are evident and manifest themselves as density undulations of the otherwise smooth (flat-top type) droplet
core.~Once again, both fragments constituting the majority component move in-sync implying that the droplet
part governs the dynamical response.~However, as $N_t$ increases, the breathing amplitude becomes more
intense and its frequency decreases, e.g., compare Figs.~\ref{fig:breathing}(e) and (g), while the majority
component fragments are less distinguishable, see Figs.~\ref{fig:dipole}(c), (d), (g), and (h).

\section{Summary and perspectives}\label{sec:conclusions}

We examined the existence and stability of two-component particle imbalanced
droplet phases in 1D as captured by the appropriate coupled set of eGPEs taking
into account the LHY quantum correction term.~In particular, the system under
investigation is a particle imbalanced homonuclear two-component bosonic mixture
featuring intracomponent repulsion and intercomponent attraction, while being weakly
confined by an external harmonic trap.~Along with the characteristics of the underlying
stationary states, we inspect the droplets' excitation spectrum and also dynamically
probe the droplets' fundamental dipole and breathing modes by utilizing suitable quenches
of the external trap.

Focusing on the ground state of the system, a multitude of complex droplet states is
identified with respect to the atom number for different imbalances and interaction
configurations.~It is found that the involved components remain miscible in nature for
the different parametric variations considered, while hosting highly deformed configurations.%
~Namely, the minority component sustains a flat-top type droplet but the majority species
exhibits mixed configurations forming a droplet in the vicinity of the minority and excess
atoms reside in a gas state at the tails of the distribution density. These mixed phases
appear to be more prominent for either increasing total atom number and stronger intercomponent
attractions while keeping the imbalance fixed or in general for larger imbalance fraction.~Naturally,
larger attractions support stronger binding among the components but lead to a smaller droplet
fragment.~In a similar vein, a fixed imbalance leads to increasingly spatially extended
distributions and favors the formation of larger droplet and excess atoms segments for increasing
atom number.

To infer the bound state character of the ensuing two-component configurations, we evaluate
the chemical potential of each component separately and the one of the entire system.~It is
explicated that the bound character of the system is highly tunable meaning that it may either
remain bound or experience a transition to a trapped gas state (mainly depending on the considered
interactions or particle imbalance).~This is traced back to the interplay of the individual
components with the minority one possessing negative chemical potential, and the majority featuring
predominantly positive chemical potentials due to the existence of excess atoms.~Overall, the
parametric regions of bound state formation are extended in terms of the total atom number for
stronger attractions or larger imbalances with the former having an arguably more substantial
effect.~As expected, smaller imbalances are associated with lower-lying composite states.{~Quite
importantly, this work demonstrates that the aforementioned two-component droplet states are
spectrally stable (in the realm of BdG analysis) over particle imbalance, atom number, and
interaction variations.}~Indeed, the spectrum contains vanishing imaginary parts, whereas its
real part connects with the droplets' collective modes.~Recall that this stability outcome is
in line with earlier predictions in the symmetric mixture case~\cite{Tylutki,katsimiga2023interactions}.

To trigger some of the characteristic droplet modes in the dynamics, we employ quenches where in
particular a trap shift induces the droplets dipole motion and a change of the trap frequency promotes
their breathing evolution.~The former manifests as a collective in-trap oscillation of the individual
clouds and the latter as a periodic expansion and contraction dynamics of the droplet distributions.%
~In both cases, the measured oscillation frequencies match with the predictions of the excitation
spectrum, i.e., the dipole mode frequency is insensitive to atom number variations and the breathing
mode one slightly increases for larger atom numbers.~Interestingly, for low atom numbers where the
minority component is substantially localized, we observe that its trajectory is imprinted in the majority
cloud as a result of back action.~In all cases, the droplet and the excess atom fragments of the majority
species oscillate in-sync.

Our findings pave the way for various fruitful extensions, aiming to advance our understanding on
multicomponent droplet settings.~In this context, it would be intriguing to examine the emergent
nonequilibrium dynamics of the imbalanced two-component droplet states, such as nucleated patterns
and the role of modulation instability~\cite{mithun2020modulational} by considering, for instance,
interaction quenches across the localized to delocalized phases of the minority atoms.~Moreover,
the generalization of our results in higher dimensions (with the appropriate modification of the
LHY term) intending to reveal the interplay of transverse modes on the droplet formation is of immense
interest.~In that vein, the computational framework developed in~\cite{SADAKA2025109378} together
with deflation-based techniques~\cite{deflation_2018,boulle2020deflation,charalampidis2020bifurcation,panos-boulle-2022} will be the principal vehicle for these higher-dimensional studies.~Also, the characterization
of strongly interacting two-component droplet phases requiring the utilization of beyond eGPE techniques
such as the ab-initio ML-MCTDHX method~\cite{cao2017unified} or exact diagonalization~\cite{Chergui_ED_drops},
for exploring the underlying correlation patterns and excitation processes is certainly desirable.~Finally,
extending our present considerations to three-component systems, especially for imbalanced mixtures which
are expected to host far richer phases and bear more complex stability properties would be worth
pursuing.~All these are exciting future research directions that are currently under consideration,
and results will be reported in forthcoming works.

\section*{Acknowledgements}
EGC is supported by the U.S. National Science Foundation under Grant No. DMS-2204782.%
~SIM acknowledges support from the Missouri Science and Technology, Department of Physics,
Startup fund.~SIM thanks P.G.~Kevrekidis, G.C.~Katsimiga, G.B.~Bougas, and J.~Pelayo for
fruitful discussions.

\appendix

\section{Spectral stability analysis of the two-component droplet setting}\label{sec:Appendix1}
{
In this Appendix, we provide the setup of the stability analysis
problem for the 1D two-component quantum droplets (see, Sec.~\ref{sec:BdG}).~The
substitution of Eq.~\eqref{pertr_ansatz_2C} into Eq.~\eqref{eq:eGPE},
and upon keeping $\mathcal{O}(\varepsilon)$ terms, gives the following
operator eigenvalue problem}
\begin{align}
-\omega
\begin{bmatrix}
a_{1}\\
b_{1}\\
a_{2}\\
b_{2}
\end{bmatrix}
=
\begin{bmatrix}
        \mathcal{L}_{11}        & \mathcal{L}_{12}  & \mathcal{L}_{13}         & \mathcal{L}_{14} \\
-\mathcal{L}_{12}^{\ast}        & -\mathcal{L}_{11} & -\mathcal{L}_{14}^{\ast} & -\mathcal{L}_{13}^{\ast} \\
        \mathcal{L}_{13}^{\ast} & \mathcal{L}_{14}  & \mathcal{L}_{33}         & \mathcal{L}_{34} \\
-\mathcal{L}_{14}^{\ast}        & -\mathcal{L}_{13} & -\mathcal{L}_{34}^{\ast} & -\mathcal{L}_{33}
\end{bmatrix}
\begin{bmatrix}
a_{1}\\
b_{1}\\
a_{2}\\
b_{2}
\end{bmatrix},
\label{eq:stab}
\end{align}

with matrix elements

%
{\small
\begin{subequations}
\begin{align}
\mathcal{L}_{11} & = -\frac{1}{2}\frac{d^{2}}{dx^{2}}%
+2\left(P+GP^{-1}\right)|\psi^{(0)}_{1}|^{2}%
-\left(1-G\right)|\psi_{2}^{(0)}|^{2}
\nonumber \\
&-\frac{P^{2}}{2\pi}\frac{|\psi_{1}^{(0)}|^{2}}%
{\sqrt{P|\psi_{1}^{(0)}|^{2}+P^{-1}|\psi_{2}^{(0)}|^{2}}}%
\nonumber \\
&-\frac{P}{\pi}\sqrt{P|\psi_{1}^{(0)}|^{2}+P^{-1}|\psi_{2}^{(0)}|^{2}}%
+V(x)-\mu_{1}, \\
\mathcal{L}_{12} & = \left[\left(P+GP^{-1}\right)-%
\frac{P^{2}}{2\pi\sqrt{P|\psi_{1}^{(0)}|^{2}+P^{-1}|\psi_{2}^{(0)}|^{2}}}\right]\left(\psi_{1}^{(0)}\right)^{2},\\
\mathcal{L}_{13} & =-\left[\left(1-G\right)%
+\frac{1}{2\pi\sqrt{P|\psi_{1}^{(0)}|^{2}+P^{-1}|\psi_{2}^{(0)}|^{2}}}\right]
\psi_{1}^{(0)}\left(\psi_{2}^{(0)}\right)^{\ast},\\
\mathcal{L}_{14} & =-\left[\left(1-G\right)%
+\frac{1}{2\pi\sqrt{P|\psi_{1}^{(0)}|^{2}+P^{-1}|\psi_{2}^{(0)}|^{2}}}\right]
\psi_{1}^{(0)}\psi_{2}^{(0)},\\
\mathcal{L}_{33} & =
-\frac{1}{2}\frac{d^{2}}{dx^{2}}+2\left(P^{-1}+GP\right)|\psi^{(0)}_{2}|^{2}%
-\left(1-G\right)|\psi_{1}^{(0)}|^{2}
\nonumber \\
&-\frac{1}{2P^{2}\pi}\frac{|\psi_{2}^{(0)}|^{2}}%
{\sqrt{P|\psi_{1}^{(0)}|^{2}+P^{-1}|\psi_{2}^{(0)}|^{2}}}
\nonumber \\
&-\frac{1}{P\pi}\sqrt{P|\psi_{1}^{(0)}|^{2}+P^{-1}|\psi_{2}^{(0)}|^{2}}%
+V(x)-\mu_{2},\\
\mathcal{L}_{34} & = \left[\left(P^{-1}+GP\right)-\frac{1}{2P^{2}\pi%
\sqrt{P|\psi_{1}^{(0)}|^{2}+P^{-1}|\psi_{2}^{(0)}|^{2}}}\right]\left(\psi_{2}^{(0)}\right)^{2}.
\end{align}
\end{subequations}
}
{This eigenvalue problem is subsequently solved numerically in order to obtain the underyling excitation spectra provided, for instance, in Fig.~\ref{fig:Bdg_8020} and Fig.~\ref{fig_Bdg_6040} of the main text.}

\bibliographystyle{apsrev4-1}
\bibliography{main.bib}

\end{document}